\documentclass[10pt,english]{IEEEtran}
\usepackage[T1]{fontenc}
\usepackage[latin9]{inputenc}
\usepackage{verbatim}
\usepackage{nomencl}

\usepackage{color}
\usepackage{float}
\usepackage{textcomp}
\usepackage{mathrsfs}
\usepackage{amsthm}
\usepackage{amsmath}
\usepackage{amssymb}
\usepackage{stackrel}
\usepackage{graphicx}
\usepackage[colorlinks]{hyperref}

\makeatletter

\floatstyle{ruled}
\newfloat{algorithm}{tbp}{loa}
\providecommand{\algorithmname}{Algorithm}
\floatname{algorithm}{\protect\algorithmname}

\floatstyle{ruled}
\newfloat{algorithm}{tbp}{loa}
\providecommand{\algorithmname}{Algorithm}
\floatname{algorithm}{\protect\algorithmname}

\newtheorem{lemma}{Lemma}

\usepackage{colortbl}
\usepackage{cite}
\usepackage{nomencl}
\usepackage{flushend}
\usepackage{stfloats}
\usepackage[caption=false]{subfig}

\@ifundefined{showcaptionsetup}{}{%
\PassOptionsToPackage{caption=false}{subfig}}
\usepackage{subfig}
\makeatother

\usepackage{babel}
\addto\captionsamerican{\renewcommand{\algorithmname}{Algorithm}}
\addto\captionsamerican{}
\addto\captionsenglish{}

\begin{document}

\title{Modeling and Design of RIS-Assisted Multi-cell Multi-band Networks with RSMA }

\author{Abdelhamid Salem, \textit{Member, IEEE}, Kai-Kit Wong, \textit{Fellow, IEEE}, Chan-Byoung Chae, \textit{Fellow, IEEE},\\and Yangyang Zhang
\thanks{A. Salem is with the Department of Electronic and Electrical Engineering, University College London, London, UK, (e-mail: a.salem@ucl.ac.uk).}
\thanks{K.-K. Wong is with the Department of Electronic and Electrical Engineering, University College London, London, UK, (e-mail: kai-kit.wong@ucl.ac.uk). He is also affiliated with Yonsei Frontier Lab, Yonsei University, Seoul, Korea.}
\thanks{C.-B. Chae is with the School of Integrated Technology, Yonsei University, Seoul, Korea, (e-mail: cbchae@yonsei.ac.kr).}
\thanks{Y. Zhang is with Kuang-Chi Science Limited, Hong Kong SAR, China (e-mail: yangyang.zhang@kuang-chi.com).}
\thanks{The work is supported by the Engineering and Physical Sciences Research Council (EPSRC) under grant EP/V052942/1 and by the National Research Foundation of Korea (NRF) Grant through the Ministry of Science and ICT (MSIT), Korea Government, under Grant 2022R1A5A1027646.}
\vspace{-8mm}
}

\maketitle
\begin{abstract}
Reconfigurable intelligent surface (RIS) has been identified as a promising technology for future wireless communication systems due to its ability to manipulate the propagation environment intelligently. RIS is a frequency-selective device, thus it can only effectively manipulate the propagation of signals within a specific frequency band. This frequency-selective characteristic can make deploying RIS in wireless cellular networks more challenging, as adjacent base stations (BSs) operate on different frequency bands. In addition, rate-splitting multiple access (RSMA) scheme has been shown to enhance the performance of RIS-aided multi-user communication systems. Accordingly, this work considers a more practical reflection model for RIS-aided RSMA communication systems, which accounts for the responses of signals across different frequency bands. To that end, new analytical expressions for the ergodic sum-rate are derived using the moment generating function (MGF) and Jensen's inequality. Based on these analytical sum-rate expressions, novel practical RIS reflection designs and power allocation strategies for the RSMA scheme are proposed and investigated to maximize the achievable sum-rate in RIS-assisted multi-cell, multi-band cellular networks. Simple sub-optimal designs are also introduced and discussed. The results validate the significant gains of our proposed reflection design algorithms with RSMA over conventional schemes in terms of achievable sum-rate. Additionally, the power allocation strategy for the RSMA scheme is shown to offer superior performance compared to conventional precoding schemes that do not rely on RSMA.
\end{abstract}


\section{Introduction}
Reconfigurable intelligent surface (RIS) technology has been proposed recently to radically change and control the wireless propagation environment \cite{RIS_chae24,RIS_dai20,Mefullduplex,Meclustering}. More specifically, RIS is a meta-surface composed of low-cost passive reflecting elements, and each element can be controlled to adjust the phase shift and amplitude of the incident signals. Due to this outstanding capability, RIS is deemed a promising and revolutionary technique to enhance the coverage and the performance of wireless communication systems \cite{RIS_chae24,RIS_dai20,Mefullduplex,Meclustering}. To leverage these advantages, extensive research on employing RIS in communication networks has been conducted. In \cite{newreff1}, the authors discussed the implementation of RIS in wireless networks and presented a communication-theoretic framework to analyze and design the RIS. New optimization problems have been formulated and solved in \cite{RefPasAct} to minimize the power consumption of  RIS-aided communication systems by optimizing the transmit active beamforming at the source and the passive beamforming at the RIS. In \cite{RefEE}, energy-efficient designs for both the RIS phase shifts and transmit power allocation have been developed. In \cite{Ref2}, the fundamental capacity limit of multiple-input multiple-output (MIMO) communication systems aided by RIS has been investigated. Further work in \cite{newreff2} presented a detailed analysis of RIS-aided single-input single-output (SISO) communication systems. Path loss models for RIS-aided wireless  systems were developed in \cite{newreff3}. Latency-minimization problems for RIS  systems were formulated and solved for single- and multi-device scenarios in \cite{Latency}. In \cite{channel3} a novel two-timescale based RIS scheme has been proposed, in which the beamforming at the base station (BS) was developed based on the
instantaneous channel state information (CSI) while the  RIS phase shifts were designed based on long-term statistical CSI. In \cite{Ref3} the
achievable ergodic rate of a RIS-aided MIMO communication systems
was derived under Rician communication channels. In \cite{Ref4},
an asymptotic ergodic sum rate for an RIS-aided communication systems
was derived based on an infinite number of BS antennas. In \cite{Ref8}
an ergodic rate expression of RIS-assisted
MIMO systems with zero forcing (ZF) beamforming scheme was presented.
Further work in \cite{Ref12} explored the average rate of RIS-MIMO communication systems under Rayleigh-Rician communication channels.
In \cite{Refsumrate,Ref13} the weighted sum-rate of RIS-aided communication systems has been maximized. The design of RIS with phase
shift errors has been explored in \cite{Refphase,RefphaseMe}.

Furthermore, rate splitting multiple access (RSMA) technique has been
introduced and investigated in various applications to enhance the
performance of multiple users multiple-input single-output (MU-MISO)
communication systems \cite{RS1}. RSMA scheme splits the users' messages
into a common and private parts which are precoded, and superimposed
in a common transmission. The users first decode the common message,
then using first layer successive interference cancellation (SIC)
scheme each private signal can be detected by the intended user. By
tuning the power allocated to the common and private parts, RSMA can
better handle the multiuser interference. Interestingly, the combination
of RIS and RSMA has attracted more attention recently. In \cite{IRSRSMA1} the potential of synergy between RIS and RSMA
has been presented, and the essential enhancements achieved by RIS-RSMA
schemes have been identified and supported via insightful results.
In \cite{IRSRSMA2} RIS and RSMA schemes have been implemented to assist users in dead-zone areas. A novel
RIS-RSMA framework to improve MU-MISO downlink
communications has been proposed in \cite{IRSRSMA3}. The work
in \cite{IRSRSMA4} considered multi-RIS-assisted RSMA systems to
provide ultra-reliable low latency communication with finite block-length
transmission. In \cite{IRSRSMA5} a downlink RSMA architecture for
RIS-aidied multi-user communication systems has been proposed. An
overview of integrating RSMA and RIS is provided in \cite{IRSRSMA6},
where the need for integrating RSMA with RIS has been explained and
a general model of a RIS-assisted RSMA system has been presented. 

Most prior work in the literature designed the optimal
RIS phase-shifts for narrow-band communication systems. However, it has been verified in several studies that the 
phase-shifts of the RIS units can vary with the frequency of incident
signals. Therefore, the conventional RIS designs, that have not considered
the impact of signal frequency, might not work properly in multi-band/wide-band
systems. Thus the design of RIS in multi-band/wide-band networks is worth studying.
The frequency-selective feature of RIS has been illustrated in some works in the literature. It has been shown that each RIS element can provide an adjustable phase-shift for messages within a specific frequency band while maintaining a nearly constant phase-shift for signals in other frequency bands \cite{wideband1,wideband2,wideband3,wideband4}. In \cite{wideband1}  a new design of the RIS in wide-band systems has been analyzed. A simpler
design of the RIS with wide-band orthogonal frequency division multiplexing
(OFDM) has been developed in \cite{wideband2}. Further work in \cite{wideband3}
provided a tractable RIS design in  multi-band systems. In \cite{wideband4} the authors optimized the
BS-RIS association taking into account  the frequency-selective
characteristics of the RIS. 

Accordingly, in this paper, we propose a simplified practical
RIS design for a RIS-assisted multi-cell multi-band RSMA
networks. More specifically, multiple BSs operate at different frequency
bands and  serve their associated users simultaneously with the aid of one RIS panel.
To enhance the network performance, each BS is proposed to implement
RSMA scheme to serve the users. Two-timescale based RIS scheme is
considered. We first derive closed-form analytical expressions for the ergodic
sum-rates using moment generating function (MGF) and Jensen inequality.
The RIS phase shifts and the service
selection of each RIS element are then designed by maximizing the
ergodic sum-rates\textcolor{blue}{.} The power allocation between
the common and private parts of RSMA scheme is also considered. This
is the first paper that studies RSMA scheme to overcome the frequency
selectivity property of RIS. Our contributions are summarized as follows:

\begin{enumerate}
\item New two closed-form analytical expressions for the ergodic
sum-rate are derived for RIS-aided multi-cell multi-band RSMA systems.
The first analytical expression is derived by using MGF, while Jensen
inequality is used to derive the second expression. In these derivations,
ZF precoding scheme is implemented for the private messages, and maximum
ratio transmission scheme (MRT) is used for the common part under
Rician fading channels. The derivation of the ergodic sum-rate of
RSMA scheme in RIS-aided communication systems under Rician channels
is challenging and hard to consider mathematically. According to the
best of our knowledge, this is the first work derives the ergodic
sum rate of RSMA scheme in RIS-aided multi-cell multi-band networks.
The obtained sum-rate formulas are explicit and offer  numerous significant practical design insights.
\item Based on the analytical sum-rate expressions, practical RIS
design is obtained by maximizing the achievable sum-rate subject to
practical RIS reflections, quality of services (Qos) and RSMA constraints.
To solve this challenging problem, fractional programming (FP) method
is used to transform the logarithmic and fractional objective function
into a more tractable form. The block coordinate descent (BCD) method
is then employed to decouple the original problem into sub-problems. Subsequently,
efficient algorithms based on the alternating direction method of multipliers
(ADMM) are developed to solve these sub-problems. A power allocation
scheme to split the total transmit power between the common and private
parts is also considered to maximize the resulting sum-rate.
\item Sub-optimal designs based on the derived expressions and QoSs are
also presented and discussed.  In the first design, we obtain the optimal RIS design that satisfies QoS of all users. In the second we consider time division (TD) protocol, in which all the RIS
elements serve only one BS in different time slots. In the third deisgn,  RIS elements division (ED) protocol is studied, where the RIS elements are divided into tiles, and each tile is assigned
to a BS.  
\item Numerical and simulation results are presented to confirm the analysis and show the effectiveness of the proposed algorithms
and confirm the substantial performance enhancement achieved by the
proposed RIS design in multi-cell multi-band communication
networks.
\end{enumerate}

\section{System Model\label{sec:System-Model}}

\begin{figure}
\noindent \begin{centering}
\includegraphics[bb=250bp 70bp 710bp 510bp,clip,scale=0.28]{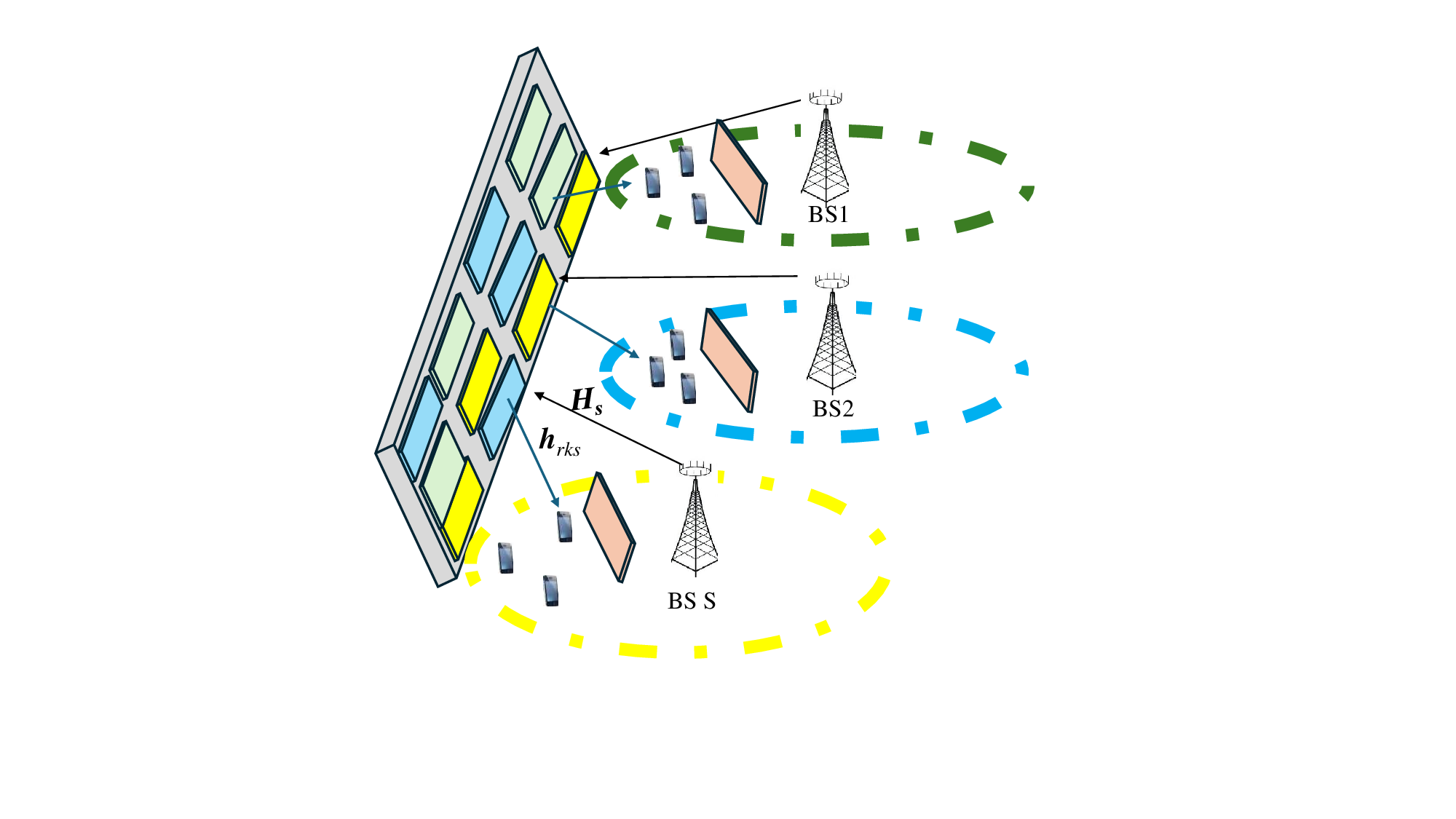}
\par\end{centering}

\protect\caption{\label{fig:model}An RIS-aided multi-cell multi-band RSMA system.}

\end{figure}

We consider a RIS-aided multi-cell multi-band RSMA network
as depicted in Fig. \ref{fig:model}, where one RIS consists of $M$
reflecting elements is implemented to assist the downlink transmissions
for $S$ BSs operating at different frequency bands.
More specific, the $s^{th}$ BS is equipped with $N_{s}$ antennas
and sends independent data to $K_{s}$ single-antenna users at
frequency $f_{s}$ using RSMA technique. The channels from the
$s^{th}$ BS to the RIS is $\mathbf{H}_{s}\in M\times N_{s}$, and from
the RIS to the $k_{s}$-th user is $\mathbf{h}_{r,k_{s}}\in M$. Considering the RIS is often installed on tall buildings and placed close to the
users, $\mathbf{H}_{s}$ is modeled as a Rician fading channel, while
$\mathbf{h}_{r,k_{s}}$ has high line-of-sight (LoS) probability \cite{channel2,channel1,channel3,channel4}. For analytical tractability,
we can write
\begin{equation}
\mathbf{H}_{s}=l_{s}\left(\sqrt{\frac{\kappa_{s}}{\kappa_{s}+1}}\mathbf{\bar{H}}_{s}+\sqrt{\frac{1}{\kappa_{s}+1}}\mathbf{\tilde{H}}_{s}\right),
\end{equation}

\noindent where $l_{s}$ is the path-loss factor, $\kappa_{s}$ is
the Rician factor, $\mathbf{\tilde{H}}_{s}$ and $\mathbf{\bar{H}}_{s}$
are the non-LoS (NLOS) and LoS components. The LoS components $\mathbf{\bar{H}}_{s}$
and $\mathbf{\bar{h}}_{r,k_{s}}$ can be expressed as 
\begin{equation}
\mathbf{\bar{H}}_{s}=\mathbf{a}_{N}\left(\phi_{r}^{a},\phi_{r}^{e}\right)\mathbf{a}_{M}^{H}\left(\phi_{t}^{a},\phi_{t}^{e}\right),\quad\mathbf{\bar{h}}_{r,k_{s}}=l_{k_{s}}\mathbf{a}_{M}\left(\phi_{ir}^{a},\phi_{ir}^{e}\right)
\end{equation}
\noindent where $l_{ks}$ is the path-loss factor, $\phi_{ir}^{a},\phi_{ir}^{e}$
are the RIS-user $i$ azimuth and elevation angles of arrival,  $\phi_{t}^{a},\phi_{t}^{e},\phi_{r}^{a},\phi_{r}^{e}$ denote the  BS-RIS azimuth
and elevation angles of departure and arrivals respectively.

\subsection{RIS Reflection Model}

Recent works illustrate that each RIS reflecting element can be adjusted
to serve a certain BS that operates with a certain frequency with
a tunable ideal phase-shift (i.e., $0\sim2\pi$ ), while serving the
other BSs that operate with different frequency bands with almost
fixed phase-shift $\left(0\textrm{ or }2\pi\right)$ \cite{wideband1,wideband2,wideband3,wideband4}.
This frequency-selective characteristic  motivates us to consider the practical RIS design using the ideal phase-shifts and a binary frequency indicator, which can
simplify the practical RIS model and reduce the complexity of RIS
design \cite{wideband3,wideband4}.

Accordingly, we can define the service selection vector for the m-th
reflecting element as $\mathbf{v}_{m}=\left[v_{1,m},...,v_{S,m}\right]^{T},\textrm{ where }v_{s,m}\in\left\{ 0,1\right\} $,
$v_{s,m}=1$ represents that $m$-th element serves the $s$-th BS
with a fully tunable phase-shift ($0\sim2\pi$) response, and $v_{s,m}=0$
indicates that the m-th reflecting element provides a fixed $\left(0\textrm{ or }2\pi\right)$
phase-shift for the $s^{th}$ BS. Thus each reflecting element can serve
only one BS, and this can be represented of element $m$ as $\left\Vert \mathbf{v}_{m}\right\Vert _{0}\leq1$,
$v_{s,m}\in\left\{ 0,1\right\} ,\forall s,m.$ Accordingly, the service
selection matrix for the RIS can be defined as $\mathbf{V}=\left[\mathbf{v}_{1},...,\mathbf{v}_{M}\right]\in S\text{\texttimes}M$.
The $s^{th}$ row in $\mathbf{V}$ represents the service selection vector
for the $s^{th}$ BS, $\mathbf{\tilde{v}}_{s}=\left[v_{s,1},...,v_{s,M}\right]^{T}$.
Thus, the service selection matrix for the RIS can also be expressed
as $\mathbf{V}=[\tilde{\mathbf{v}}_{1},...,\tilde{\mathbf{v}}_{S}]^{T}$.
In addition, the ideal phase-shift vector for the $s^{th}$ BS is $\mathbf{\phi}_{s}=\left[\phi_{s,1},...,\phi_{s,M}\right]^{T},\phi_{s,m}\in(0,2\pi]$.
Finally, combining both the service selection and the ideal phase
shift, we can introduce the practical RIS reflection model of the
$s^{th}$ BS as $\mathbf{\theta}_{s}=e^{j\mathbf{\phi}_{s}\odot\mathbf{\tilde{v}}_{s}},\forall s$,
$\phi_{s,m}\in(0,2]$, $\left\Vert \mathbf{v}_{m}\right\Vert _{0}\leq1$,
$v_{s,m}\in\left\{ 0,1\right\} ,\forall s,m$, and the reflecting
coefficient matrix of the RIS as $\Theta_{s}=\textrm{diag}\left\{ \mathbf{\theta}_{s}\right\} \textrm{where }\mathbf{\theta}_{s}=\left[\theta_{s,1},...,\theta_{s,M}\right]^{T}$.

\subsection{Rate Splitting Signal Model}

By implementing RSMA scheme, each BS splits each  message into
private and common parts. For the $s^{th}$ BS the signal
of the $k_{s}^{th}$ user is split into common part, $x_{s,c_{k_{s}}}$
and private part, $x_{s,k_{s}}$. The common parts of all users
will be packed to form the common message, $x_{s,c}=\left\{ x_{s,c_{1}},...,x_{s,c_{K_{s}}}\right\} $.
The resulting $K_{s}+1$ symbols can be presented as $\mathbf{x}_{s}=\left[x_{s,c},x_{s,1},....,x_{s,K_{s}}\right]^{T}\in\mathbb{C}^{K_{s}+1}$,
where $\mathcal{E}\left\{ \mathbf{x}_{s}\mathbf{x}_{s}^{H}\right\} =\textrm{I}$.
The symbols are then precoded via a precoding matrix defined
by $\mathbf{W}_{s}=\left[\mathbf{w}_{s,c},\mathbf{w}_{s,1},....\mathbf{w}_{s,K_{s}}\right]$
where $\mathbf{w}_{s,c}\in\mathbb{C}^{N_{s}}$ is the common precoder
vector and $\mathbf{w}_{s,k_{s}}\in\mathbb{C}^{N_{s}}$ is the $k_{s}^{th}$
private precoder vector. Accordingly, the transmitted signal by the
$s^{th}$ BS is given by \cite{RS1}
\begin{equation}
\mathbf{s}_{s}=\mathbf{W}_{s}\mathbf{x}_{s}=\sqrt{P_{s,c}}\mathbf{w}_{s,c}x_{s,c}+\stackrel[i=1]{K_{s}}{\sum}\sqrt{P_{s,p}}\mathbf{w}_{s,i}x_{s,i},
\end{equation}

\noindent where $P_{s,c}=\left(1-t_{s}\right)P_{s}$ is the power
allocated to the common message at the $s^{th}$ BS with $0<t_{s}\leq1$
and $P_{s}$ is the total power, while $P_{s,p}=\dfrac{t_{s}P_{s}}{K_{s}}$
is the power allocated to the private message at the $s^{th}$ BS.
The received signal at the $k_{s}$-th user can be expressed as
\begin{equation}
y_{s,k_{s}}=\sqrt{P_{s,c}}\mathbf{g}_{k_{s}}\mathbf{w}_{s,c}x_{s,c}+\mathbf{g}_{k_{s}}\stackrel[i=1]{K_{s}}{\sum}\sqrt{P_{s,p}}\mathbf{w}_{s,i}x_{s,i}+n_{k_{s}},\label{eq:3-1}
\end{equation}

\noindent where $\mathbf{g}_{k_{s}}=\mathbf{h}_{r,k_{s}}^{H}\Theta_{s}\mathbf{H}_{s}$,
$n_{k_{s}}\sim\mathcal{CN}\left(\text{0, }\sigma_{k_{s}}^{2}\right)$
is the additive \textit{\emph{white}} Gaussian noise (AWGN). At reception,
each user first detects the common message and then remove it using
SIC technique. Consequently, the received private message at the $k^{th}$
user is given by
\begin{eqnarray}
y_{s,k_{s}}^{p} & == & \mathbf{g}_{k_{s}}\stackrel[i=1]{K_{s}}{\sum}\sqrt{P_{s,p}}\mathbf{w}_{s,i}x_{s,i}+n_{k_{s}},\label{eq:4}
\end{eqnarray}

The signal to interference and noise ratio (SINR) of the common and
private messages can be expressed as

\begin{equation}
\gamma_{s,k_{s}}^{c}=\frac{P_{s,c}\left|\mathbf{g}_{k_{s}}\mathbf{w}_{s,c}\right|^{2}}{P_{s,p}\stackrel[i=1]{K_{s}}{\sum}\left|\mathbf{g}_{k_{s}}\mathbf{w}_{s,i}\right|^{2}+\sigma_{k_{s}}^{2}}.\label{eq:7}
\end{equation}
and
\begin{equation}
\gamma_{s,k_{s}}^{p}=\frac{P_{s,p}\left|\mathbf{g}_{k_{s}}\mathbf{w}_{s,k_{s}}\right|^{2}}{P_{s,p}\stackrel[\underset{i\neq k_{s}}{i=1}]{K_{s}}{\sum}\left|\mathbf{g}_{k_{s}}\mathbf{w}_{s,i}\right|^{2}+\sigma_{k_{s}}^{2}}.\label{eq:8}
\end{equation}

\noindent The sum rate of RSMA scheme for the $s^{th}$ BS can be
calculated by
\begin{equation}
R_{s}=R_{s,c}+\stackrel[k=1]{K_{s}}{\sum}R_{s,k}^{p},\label{eq:5}
\end{equation}

\noindent where $R_{s,c}=\min\left(R_{s,1}^{c},..,R_{s,K_{s}}^{c}\right)$, $R_{s,k}^{c}$ is the common rate at the
$k_{s}^{th}$ user, and $R_{s,k}^{p}$ is the rate of the private
message. The channel coding is performed over
a long channel status, thus transmitting the common signal and the
private signals s at ergodic rates $\min_{j}\left(\mathcal{E}\left\{ R_{s,j}^{c}\right\} \right)_{j=1}^{K_{s}}$
and $\mathcal{E}\left\{ R_{s,k}^{p}\right\} $, respectively, can
guarantee successful detection at the user. Consequently, the ergodic
sum rate of the $s^{th}$ BS can be expressed as
\begin{equation}
\mathcal{E}\left\{ R_{s}\right\} =\min_{j}\left(\mathcal{E}\left\{ R_{s,j}^{c}\right\} \right)_{j=1}^{K_{s}}+\stackrel[k=1]{K_{s}}{\sum}\mathcal{E}\left\{ R_{s,k}^{p}\right\} \label{eq:5-1}
\end{equation}

The ergodic sum-rate has been analyzed in the next section  using different  approaches.

\section{Ergodic Sum-Rate Analysis\label{sec:Ergodic-Rates-Analysis}}

Here we analyze the ergodic sum-rate for each BS in the
network. For mathematical tractability but without loss of generality,
closed form precoding techniques are implemented for the common and
private messages. Precisely, the common message won't cause any interference
in decoding the private signals due to use SIC. Thus, MRT precoding
scheme is implemented for the common message. Also, to mitigate
the inter-user interference ZF precoding technique
is used for the private messages. Note that such precoding design
has been adopted and validated in the literature \cite{Ref8,channel2}.
Thus, the MRT and the ZF precoders can be presented, respectively,
as

\begin{equation}
\mathbf{w}_{s,c}=\frac{\mathbf{g}_{k_{s}}^{H}}{\left\Vert \mathbf{g}_{k_{s}}^{H}\right\Vert },\,\mathbf{w}_{s,k_{s}}=\frac{\mathbf{f}_{k_{s}}}{\left\Vert \mathbf{f}_{k_{s}}\right\Vert }\label{eq:11}
\end{equation}

\noindent where $\mathbf{f}_{k_{s}}$ is the $k_{s}$ th column of
the matrix $\mathbf{F}_{s}=\mathbf{G}_{s}^{H}\left(\mathbf{G}_{s}\mathbf{G}_{s}^{H}\right)^{-1}$,
$\mathbf{G}_{s}=\mathbf{H}_{r,s}^{H}\Theta_{s}\mathbf{H}_{s}$ ,$\mathbf{H}_{r,s}^{H}$
is the channel matrix between the RIS and the $K_{s}$ users.

\subsection{MGF Approach}

By plugging the precdoing vectors in (\ref{eq:11}) into (\ref{eq:7})
the common SINR can be re-presented as

\begin{equation}
\gamma_{s,k_{s}}^{c}=\frac{P_{s,c}\left\Vert \mathbf{g}_{k_{s}}\right\Vert ^{2}}{\frac{P_{s,p}}{\left[\left(\mathbf{G}_{s}\mathbf{G}_{s}^{H}\right)^{-1}\right]_{k_{s},k_{s}}}+\sigma_{k_{s}}^{2}}.\label{eq:13}
\end{equation}

\noindent The ergodic  rate of user $k_{s}$ common message is
 then expressed as
\begin{equation}
\mathscr{E}\left[R_{s,k_{s}}^{c}\right]=\mathscr{E}\left[\log_{2}\left(1+\gamma_{s,k_{s}}^{c}\right)\right].
\end{equation}

\begin{lemma}
The ergodic rate of the common part at the $k_{s}^{th}$ user associated
with the $s^{th}$ BS can be calculated by
\[
\mathscr{E}\left[R_{s,k_{s}}^{c}\right]=\frac{1}{\ln\left(2\right)}\,\stackrel[i=1]{n}{\sum}\textrm{H}_{i}\frac{1}{z_{i}}\left(1-\left(1+\frac{2\nu P_{s,c}z_{i}}{\sigma_{k_{s}}^{2}}\right)^{\frac{-N_{s}}{2}}\right)
\]

\begin{equation}
\times\left(1+\frac{P_{s,p}z_{i}}{\sigma_{k_{s}}^{2}\Psi_{k}}\right)^{K_{s}-N_{s}-1}\label{eq:15}
\end{equation}

\noindent where $\nu=\left(\frac{1}{1+\kappa_{s}}\mathbf{h}_{r,k_{s}}^{H}\mathbf{h}_{r,k_{s}}+\frac{\kappa_{s}}{1+\kappa_{s}}\mathbf{h}_{r,k_{s}}^{H}\Theta_{s}\bar{\mathbf{H}}_{s}^{H}\bar{\mathbf{H}}_{s}\Theta_{s}^{H}\mathbf{h}_{r,k_{s}}\right)$,
$\Psi_{k}=\left[\left(\left(\frac{1}{1+\kappa_{s}}\right)\mathbf{H}_{r}^{H}\mathbf{H}_{r}+\left(\frac{\kappa_{s}}{1+\kappa_{s}}\right)\mathbf{H}_{r}^{H}\Theta_{s}\mathbf{a}_{M}^{H}\mathbf{a}_{M}\Theta_{s}^{H}\mathbf{H}_{r}\right)^{-1}\right]_{k_{s},k_{s}},$
$z_i$ and $\textrm{H}_{i}$ are the $i^{th}$ zero and the weighting factors,
respectively, of the Laguerre polynomials tabulated in \cite[(25.245)]{book}. \end{lemma}
\begin{IEEEproof}
\noindent The proof is presented in Appendix A.
\end{IEEEproof}
By substituting the private precdoing vector in (\ref{eq:11}) into
(\ref{eq:8}) the private SINR can be re-presented as

\begin{equation}
\gamma_{s,k_{s}}^{p}=\frac{P_{s,p}}{\sigma_{k_{s}}^{2}\left[\left(\mathbf{G}_{s}\mathbf{G}_{s}^{H}\right)^{-1}\right]_{k_{s},k_{s}}}.\label{eq:8-1}
\end{equation}

\noindent The ergodic  rate of user $k_{s}$ private  message is
 expressed as

\begin{equation}
\mathscr{E}\left[R_{s,k_{s}}^{p}\right]=\mathscr{E}\left[\log_{2}\left(1+\gamma_{s,k_{s}}^{p}\right)\right].
\end{equation}

\begin{lemma}
The ergodic rate of the private message at the $k_{s}^{th}$ user
associated with the $s^{th}$ BS can be calculated by

\[
\mathscr{E}\left[R_{s,k_{s}}^{p}\right]=\stackrel[i=1]{n}{\sum}\textrm{H}_{i}\frac{1}{y_{i}}\log_{2}\left(1+\frac{P_{s,p}y_{i}}{\sigma_{k_{s}}^{2}}\right)
\]

\begin{equation}
\times\frac{\left(\Psi_{k}\right)^{N_{s}-K_{s}+1}\, y_{i}^{\left(N_{s}-K\right)}e^{-\Psi_{k}y_{i}}}{\Gamma\left(N_{s}-K_{s}+1\right)}dy\label{eq:19}
\end{equation}

\end{lemma}
\begin{IEEEproof}
\noindent The proof is presented in Appendix A.

Finally by substituting (\ref{eq:15}) and (\ref{eq:19}) into (\ref{eq:5-1})
we obtain the ergodic sum-rate.
\end{IEEEproof}

\subsection{Approximation Using Jensen inequality}

A simpler ergodic sum-rate expression is derived here based on Jensen
inequality. Thus, an approximation of the ergodic rate of the common
message can be calculated by 

\begin{equation}
\mathscr{E}\left[R_{s,k_{s}}^{c}\right]\approx\log_{2}\left(1+\mathscr{E}\left\{ \gamma_{s,k_{s}}^{c}\right\} \right).
\end{equation}

\textcolor{blue}{}
\begin{lemma}
The ergodic rate of the common message at the $k_{s}^{th}$ user associated
with the $s^{th}$ BS can be approximated by

\[
\mathscr{E}\left[R_{s,k_{s}}^{c}\right]\approx\log_{2}\left(1+\right.
\]

\begin{equation}
\left.\frac{\frac{P_{s,c}}{\left(\kappa_{s}+1\right)^{2}}\left(\kappa_{s}^{2}a_{1}+2\kappa_{s}N_{s}M\left|f_{k}\right|^{2}+N_{s}M^{2}\left(N_{s}+1\right)+2\kappa_{s}a_{2}\right)}{P_{s,p}\frac{1}{\left[\frac{1}{N_{s}-K_{s}-1}\left(a_{3}\right)^{-1}\right]_{k_{s},k_{s}}}+\sigma_{k_{s}}^{2}}\right)
\end{equation}

\noindent where $a_{1}=\left|\mathbf{\bar{h}}_{r,k}^{H}\Theta_{s}\mathbf{\bar{H}}\mathbf{\bar{H}}^{H}\Theta_{s}^{H}\mathbf{\bar{h}}_{r,k}\right|^{2}$,
$f_{k}=\mathbf{a}_{N}\left(\phi_{r}^{a},\phi_{r}^{e}\right)\Theta_{s}\mathbf{\bar{h}}_{r,k}^{H}$,
$a_{2}=\mathbf{\bar{h}}_{r,k}^{H}\Theta_{s}\mathbf{\bar{H}}\mathbf{\bar{H}}^{H}\Theta_{s}\mathbf{\bar{h}}_{r,k}\mathbf{\bar{h}}_{r,k}^{H}N\mathbf{I}_{M}\mathbf{\bar{h}}_{r,k}$
and $a_{3}=\left(\frac{1}{1+\kappa_{s}}\right)\mathbf{H}_{r}^{H}\mathbf{H}_{r}+\left(\frac{\kappa_{s}}{1+\kappa_{s}}\right)\mathbf{H}_{r}^{H}\Theta_{s}\mathbf{a}_{M}^{H}\mathbf{a}_{M}\Theta_{s}^{H}\mathbf{H}_{r}$.\end{lemma}
\begin{IEEEproof}
\noindent The proof is presented in Appendix B.
\end{IEEEproof}
Now, an approximation of the ergodic rate of the private 
message using Jensen inequality can be calculated by  
\begin{equation}
\mathscr{E}\left[R_{s,k_{s}}^{p}\right]\approx\log_{2}\left(1+\mathscr{E}\left\{ \frac{P_{s,p}}{\sigma_{k_{s}}^{2}\left[\left(\mathbf{G}_{s}\mathbf{G}_{s}^{H}\right)^{-1}\right]_{k_{s},k_{s}}}\right\} \right)\label{eq:19-1}
\end{equation}

\begin{lemma}
The ergodic rate of the private message at the $k_{s}^{th}$ user
associated with the $s^{th}$ BS can be approximated by
\end{lemma}
\begin{equation}
\mathscr{E}\left[R_{s,k_{s}}^{p}\right]\approx\log_{2}\left(1+\frac{P_{s,p}}{\left[\frac{1}{N_{s}-K_{s}-1}\left(a_{3}\right)^{-1}\right]_{k_{s},k_{s}}\sigma_{k_{s}}^{2}}\right)\label{eq:22}
\end{equation}
The proof of (\ref{eq:22}) is based on the derivation in Appendix B.

\subsection{Design Insights}

The ergodic sum-rate expressions using the MGF and Jensen inequality
presented in Lemmas 1-4 produce very accurate results. The accuracy
of the above derivations is also confirmed  by its
match with simulated results in our numerical results section. These
analytical expressions can be used to design the optimal system parameters
and obtain the optimal performance. From the sum rate expressions,
we can observe the following insights.  Generally, the practical
phase shifts, $\Theta_{s}$, which include the ideal phase shifts,
$\mathbf{\phi}_{s}$, and service selection matrix, $\mathbf{V}$,
should be optimized to enhance the system performance of each BS.
In addition, the performance of the RIS is highly related to the power
transmission, BS antennas, and number of the RIS elements, where increasing
the value of these parameters always enhances the achievable sum-rate.
In case the RIS has a limited number of reflecting elements, the performance
of each BS can be enhanced by increasing the transmit power and/or
number of BS antennas. The RIS location has an essential impact on
the system performance, as the distance to the BSs and users increases
the achievable sum-rate degrades due to larger path-loss. In line
with the basic principles of RSMA scheme, optimal power allocation
between the common and private parts should be considered carefully.
The optimal value of the power fraction for each BS, $t_{s}$, that
makes RSMA outperform the conventional schemes should be obtained
for each system setting.
Therefore, the achievable sum-rate can be enhanced by optimizing
the ideal RIS phase shifts, the service selection matrix, and the
power fraction for given system settings. This will be investigated in the following Sections.

\section{Sum-Rate Maximization Problem \label{sec:Maximization-Problem}}

In this section, we aim to maximize the achievable sum-rates of the
BSs by optimizing the ideal RIS phase-shifts, the service selection
matrix and the power between the common and private
messages, subject to QoS, RSMA and practical RIS reflection constraints.
Therefore, the sum-rate maximization problem can be formulated as
\begin{eqnarray}
\underset{\mathbf{V},\Phi,t_{s}}{\max}\,\underset{s\in S}{\sum}\hat{R_{s}}\qquad\qquad\qquad\nonumber \\
\mathbf{s.t.}\,\mathbf{\theta}_{s}=e^{j\mathbf{\phi}_{s}\odot\mathbf{\tilde{v}}_{s}},\:\forall S\qquad\qquad & (C1)\nonumber \\
\phi_{s,m}\in(0,2\pi],\:\forall S,m\qquad\qquad & (C2)\nonumber \\
\left\Vert \mathbf{v}_{m}\right\Vert _{0}\leq1,v_{s,m}\in\left\{ 0,1\right\} ,\:\forall s,m & (C3)\nonumber \\
R_{s,k}\geq R_{s,k_{s}}^{th},\:\forall s,k_{s}\qquad\qquad\qquad & (C4)\nonumber \\
R_{c,k_{s}}\geq R_{c,s},\forall s,k_{s},\qquad\qquad\qquad & (C5)\nonumber \\
0<t_{s}\leq1,,\forall s\qquad\qquad\qquad & (C6)\nonumber \\
p_{s,c}+\underset{k_{s}\in K_{s}}{\sum}p_{s,k_{s}}\leq P_{s},\:\forall s,k_{s}. & (C7)\label{eq:23}
\end{eqnarray}

\noindent where $R_{s,k_{s}}^{th}>0$ is the minimum rate requirement
of the $k_{s}$-th user served by the $s$-th BS.  The first constraints
(C1)-(C3) are for the RIS design, while (C4) illustrates that the
achievable rate of the user should be larger than the threshold, which
satisfies its minimum QoS requirements and guarantees the user fairness.
The constraint (C5) guarantees a successful decoding of the common
message, and (C6)-(C7) to ensure that the transmit power is less than
the available power budget. 

\noindent Addressing this non-convex problem is challenging due to the complexity of the objective function, the variables involved, and the constraints. To overcome these challenges, we employ the BCD method to decompose the original problem into various sub-problems, where the design of the optimal phase shift and service selection at the RIS and the power allocation are handled independently. 

Based on (\ref{eq:23}), the ideal phase shifts and service selection
matrix can be obtained by considering the following problem 
\begin{eqnarray}
\underset{\Phi,\mathbf{V}}{\max}\,\underset{s\in S}{\sum}\hat{R_{s}}\qquad\qquad\qquad\nonumber \\
\mathbf{s.t.}\,\mathbf{\theta}_{s}=e^{j\mathbf{\phi}_{s}\odot\mathbf{\tilde{v}}_{s}},\:\forall S\qquad\qquad\nonumber \\
\phi_{s,m}\in(0,2\pi],\:\forall S,m\qquad\qquad\nonumber \\
\left\Vert \mathbf{v}_{m}\right\Vert _{0}\leq1,v_{s,m}\in\left\{ 0,1\right\} ,\:\forall s,m\nonumber \\
R_{s,k_{s}}\geq R_{s,k_{s}}^{th},\:\forall s,k_{s}\qquad\qquad\qquad\label{eq:24}
\end{eqnarray}

Here, we provide a simple and very efficient method to solve this
problem. To enhance the assessment of each BS proficiency in leveraging
the RIS and to determine service selection. We initially assume an
ideal RIS for each BS and optimal phase-shift configurations for each
BS are obtained. Then, we derive the optimal service selection matrix
by allocating the suitable reflecting elements to each BS.

\subsection{Ideal Phase-shift Design \label{sub:Ideal-Phase-shift}}

As we have explained above, here we assume the RIS is ideal for all
BSs, i.e., $\tilde{\mathbf{v}}_{s}=1,\forall s$, and ideal phase-shifts
are designed for each BS. Therefore, the problem of the $s^{th}$
BS can be reformulated as
\begin{eqnarray}
\underset{\mathbf{\phi}_{s}}{\max}\,\stackrel[k_{s}=1]{K_{s}}{\sum}R_{s,k_{s}}^{p}\qquad\qquad\qquad\nonumber \\
\mathbf{s.t.}\,\phi_{s,m}\in(0,2\pi],\:\forall m\qquad\qquad\nonumber \\
R_{s,k}\geq R_{s,k}^{th},\:\forall k_{s}\qquad\qquad\label{eq:25}
\end{eqnarray}

To obtain the ideal phase shifts, we apply the FP method to simplify the logarithmic and fractional objective function.
 To start with, we first represent $R_{s,k_{s}}^{p}$
in (\ref{eq:22}) as
\begin{equation}
R_{s,k_{s}}^{p}=\log_{2}\left(1+\frac{\left(N-K-1\right)P_{s,p}\mathbf{\theta_{s}}^{H}\mathbf{B}_{ks}\mathbf{\theta_{s}}}{\theta_{s}^{H}\mathbf{A}_{ks}\mathbf{\theta_{s}}\sigma_{k_{s}}^{2}}\right)
\end{equation}

\noindent where $\mathbf{B}_{ks}=\frac{1}{M}\mathbf{I}_{M}+\delta\textrm{diag}\left(\mathbf{a}_{M}\right)\mathbf{H}_{r}\varLambda^{-1}\mathbf{H}_{r}^{H}\textrm{diag}\left(\mathbf{a}_{M}^{H}\right)$
and $\mathbf{A}_{ks}=\left\{ \left[\varLambda^{-1}\right]_{k_{s},k_{s}}\mathbf{B}_{ks}-\delta\mathbf{s}\mathbf{s}^{H}\right\} $,
$\delta=\left(\frac{\kappa_{s}}{1+\kappa_{s}}\right)$ and $\mathbf{s}=\left[\varLambda^{-1}\mathbf{H}_{r}^{H}\textrm{diag}\left(\mathbf{a}_{M}^{H}\right)\right]_{\left(:,k\right)}$.
\begin{IEEEproof}
\noindent The proof is presented in Appendix C.
\end{IEEEproof}
Now, we can rewrite (\ref{eq:25}) as 

\begin{eqnarray}
\underset{\mathbf{\phi}_{s}}{\max}\,\stackrel[k_{s}=1]{K_{s}}{\sum}\log_{2}\left(1+\frac{\left(N-K-1\right)P_{s,p}\mathbf{\theta_{s}}^{H}\mathbf{B}_{ks}\mathbf{\theta_{s}}}{\theta_{s}^{H}\mathbf{A}_{ks}\mathbf{\theta_{s}}\sigma_{k_{s}}^{2}}\right)\nonumber \\
\mathbf{s.t.}\,\phi_{s,m}\in(0,2\pi],\:\forall m\qquad\qquad\nonumber \\
\frac{\mathbf{\theta_{s}}^{H}\mathbf{B}_{ks}\mathbf{\theta_{s}}}{\theta_{s}^{H}\mathbf{A}_{ks}\mathbf{\theta_{s}}}\geq\Gamma_{s,k_{s}}^{th},\:\forall k_{s}\qquad\qquad\label{eq:27}
\end{eqnarray}

\noindent where $\Gamma_{s,k_{s}}^{th}=2^{R_{s,k}^{th}}-1$. To apply
FP method, we introduce the dual variable $\alpha=\left[\alpha_{1,1},...,\alpha_{S,K_{s}}\right]^{T}$,
thus the problem can be equivalently written as
\begin{eqnarray}
\underset{\mathbf{\phi}_{s},\alpha}{\max}\, f_{1}\qquad\qquad\qquad\nonumber \\
\mathbf{s.t.}\,\phi_{s,m}\in(0,2\pi],\:\forall m\qquad\qquad\nonumber \\
\frac{\mathbf{\theta_{s}}^{H}\mathbf{B}_{ks}\mathbf{\theta_{s}}}{\theta_{s}^{H}\mathbf{A}_{ks}\mathbf{\theta_{s}}}\geq\Gamma_{s,k_{s}}^{th},\:\forall k_{s}\qquad\qquad
\end{eqnarray}

\noindent where $f_{1}=\stackrel[k_{s}=1]{K_{s}}{\sum}\log_{2}\left(1+\alpha_{ks}\right)-\stackrel[k_{s}=1]{K_{s}}{\sum}\alpha_{ks}+\stackrel[k_{s}=1]{K_{s}}{\sum}\frac{\left(1+\alpha_{ks}\right)\mathbf{\theta_{s}}^{H}\mathbf{B}_{ks}\mathbf{\theta_{s}}}{\mathbf{\theta_{s}}^{H}\mathbf{C}_{ks}\mathbf{\theta_{s}}}$
and $\mathbf{C}_{ks}=\mathbf{A}_{ks}+\mathbf{B}_{ks}$. For a given
$\alpha$, we can focus on the following problem 
\begin{eqnarray}
\underset{\mathbf{\phi}_{s},\alpha}{\max}\, f_{2}\qquad\qquad\qquad\nonumber \\
\mathbf{s.t.}\,\phi_{s,m}\in(0,2\pi],\:\forall m\nonumber \\
\mathbf{\theta_{s}}^{H}\mathbf{B}_{ks}\mathbf{\theta_{s}}-\Gamma_{s,k_{s}}^{th}\theta_{s}^{H}\mathbf{A}_{ks}\mathbf{\theta_{s}}\geq0,\:\forall k_{s}
\end{eqnarray}

\noindent where $f_{2}=\stackrel[k_{s}=1]{K_{s}}{\sum}\frac{\left(1+\alpha_{ks}\right)\mathbf{\theta_{s}}^{H}\mathbf{B}_{ks}\mathbf{\theta_{s}}}{\mathbf{\theta_{s}}^{H}\mathbf{C}_{ks}\mathbf{\theta_{s}}}$.\textcolor{red}{{}
}The fraction $f_{2}$ can be also expressed as $f_{2}=\stackrel[k_{s}=1]{K_{s}}{\sum}\left(1+\alpha_{ks}\right)\mathbf{\theta_{s}}^{H}\mathbf{\bar{B}}_{ks}\left(\mathbf{\theta_{s}}^{H}\mathbf{C}_{ks}\mathbf{\theta_{s}}\right)^{-1}\mathbf{\bar{B}}_{ks}\mathbf{\theta_{s}}$
where $\mathbf{\bar{B}}_{ks}=\mathbf{B}_{ks}^{1/2}$. The problem
now is sum of ratios, thus by applying the quadratic transform in
\cite{FP1,FP2} $f_{2}$ can be further presented as , $f_{3}=\stackrel[k_{s}=1]{K_{s}}{\sum}2\sqrt{\left(1+\alpha_{ks}\right)}\mathbf{Re}\left\{ \mathbf{\theta_{s}}^{H}\mathbf{\bar{B}}_{ks}\mathbf{y}_{m}\right\} -\stackrel[k_{s}=1]{K_{s}}{\sum}\mathbf{y}_{m}^{T}\left(\mathbf{\theta_{s}}^{H}\mathbf{C}_{ks}\mathbf{\theta_{s}}\right)\mathbf{y}_{m}$,
where $\mathbf{y}_{m}$ refers to a collection of auxiliary variables
${y_{1},...,y_{M}}$. Thus, the problem can be re-formulated as 
\begin{eqnarray}
\underset{\mathbf{\phi}_{s},\alpha,\mathbf{y}}{\max}\,\stackrel[k_{s}=1]{K_{s}}{\sum}2\sqrt{\left(1+\alpha_{ks}\right)}\mathbf{Re}\left\{ \mathbf{\theta_{s}}^{H}\mathbf{\bar{B}}_{ks}\mathbf{y}_{K_{s}}\right\} \nonumber \\
-\stackrel[k_{s}=1]{K_{s}}{\sum}\mathbf{y}_{k_{s}}^{T}\left(\mathbf{\theta_{s}}^{H}\mathbf{C}_{ks}\mathbf{\theta_{s}}\right)\mathbf{y}_{k_{s}}\nonumber \\
\mathbf{s.t.}\,\phi_{s,m}\in(0,2\pi],\:\forall m\nonumber \\
\mathbf{\theta_{s}}^{H}\mathbf{B}_{ks}\mathbf{\theta_{s}}-\Gamma_{s,k_{s}}^{th}\theta_{s}^{H}\mathbf{A}_{ks}\mathbf{\theta_{s}}\geq0,\:\forall k_{s}
\end{eqnarray}

Now we can adopt the BCD method to handle the optimization variables,
$\alpha$, $\mathbf{y}_{m}$ and $\mathbf{\phi}_{s}$. 

\emph{Update }$\alpha$, and $\mathbf{y}_{m}$: with fixed $\mathbf{y}_{m}$
and $\mathbf{\phi}_{s}$, the optimal solution of $\alpha$ is found
to be \cite{FP1,FP2} $\alpha_{ks}^{*}=\frac{\mathbf{\theta_{s}}^{H}\mathbf{B}_{ks}\mathbf{\theta_{s}}}{\theta_{s}^{H}\mathbf{A}_{ks}\mathbf{\theta_{s}}}$.
Similarly, the optimal solution of $\mathbf{y}_{m}$ is $\mathbf{y}_{m}^{*}=\left(\mathbf{\theta_{s}}^{H}\mathbf{C}_{ks}\mathbf{\theta_{s}}\right)^{-1}\mathbf{\bar{B}}_{ks}\mathbf{\theta_{s}}$. 

\emph{Update Ideal Passive shifts $\mathbf{\theta_{s}}$: }with fixed, $\alpha$, $\mathbf{y}_{m}$, the problem can be reformulated as 
\begin{eqnarray}
\underset{\mathbf{\theta_{s}}}{\max}\,\stackrel[k_{s}=1]{K_{s}}{\sum}2\sqrt{\left(1+\alpha_{ks}\right)}\mathbf{Re}\left\{ \mathbf{\theta_{s}}^{H}\mathbf{\bar{B}}_{ks}\mathbf{y}_{m}\right\} \nonumber \\
-\mathbf{y}_{m}^{T}\left(\mathbf{\theta_{s}}^{H}\mathbf{C}_{ks}\mathbf{\theta_{s}}\right)\mathbf{y}_{m}\nonumber \\
\mathbf{s.t.}\,\left|\mathbf{\theta}_{s,m}\right|=1\forall m\nonumber \\
\mathbf{\theta_{s}}^{H}\mathbf{B}_{ks}\mathbf{\theta_{s}}-\Gamma_{s,k_{s}}^{th}\theta_{s}^{H}\mathbf{A}_{ks}\mathbf{\theta_{s}}\geq0,\:\forall s,k_{s}
\end{eqnarray}

\noindent The problem is still hard and difficult to solve due to
the non-convexity of the unit modulus constraint. ADMM method is employed
to handle this problem, thus we introduce the auxiliary variables $\mathbf{\psi}_{s}=\left[\mathbf{\psi}_{s,1},...,\mathbf{\psi}_{s,M}\right]^{T}$
and reformulate the problem as

\begin{eqnarray}
\underset{\mathbf{\theta_{s}},\mathbf{\psi}_{s}}{\max}\,\stackrel[k_{s}=1]{K_{s}}{\sum}2\sqrt{\left(1+\alpha_{ks}\right)}\mathbf{Re}\left\{ \mathbf{\theta_{s}}^{H}\mathbf{\bar{B}}_{ks}\mathbf{y}_{m}\right\} \nonumber \\
-\stackrel[k_{s}=1]{K_{s}}{\sum}\mathbf{y}_{m}^{T}\left(\mathbf{\theta_{s}}^{H}\mathbf{C}_{ks}\mathbf{\theta_{s}}\right)\mathbf{y}_{m}\nonumber \\
\mathbf{s.t.}\,\left|\mathbf{\theta}_{s,m}\right|\leq1\forall m\nonumber \\
\left|\mathbf{\psi}_{s,m}\right|=1\forall m\nonumber \\
\mathbf{\theta_{s}}=\mathbf{\psi}_{s}\forall m\nonumber \\
\mathbf{\theta_{s}}^{H}\mathbf{B}_{ks}\mathbf{\theta_{s}}-\Gamma_{s,k_{s}}^{th}\theta_{s}^{H}\mathbf{A}_{ks}\mathbf{\theta_{s}}\geq0,\:\forall s,k_{s}
\end{eqnarray}

The optimal solution can be obtained by solving the augmented Lagrangian (AL) problem. By defining the dual variables $\mathbf{\zeta}_{s}=\left[\mathbf{\zeta}_{s,1},...,\mathbf{\zeta}_{s,M}\right]^{T}$
and penalty coefficient $\rho>0$, the problem can be expressed as
\begin{eqnarray}
\underset{\mathbf{\theta_{s}},\mathbf{\psi}_{s},\mathbf{\zeta}_{s}}{\max}\,\stackrel[k_{s}=1]{K_{s}}{\sum}2\sqrt{\left(1+\alpha_{ks}\right)}\mathbf{Re}\left\{ \mathbf{\theta_{s}}^{H}\mathbf{\bar{B}}_{ks}\mathbf{y}_{m}\right\} \nonumber \\
-\stackrel[k_{s}=1]{K_{s}}{\sum}\mathbf{y}_{m}^{T}\left(\mathbf{\theta_{s}}^{H}\mathbf{C}_{ks}\mathbf{\theta_{s}}\right)\mathbf{y}_{m}\nonumber \\
-\mathbf{Re}\left\{ \mathbf{\zeta}_{s}^{H}\left(\mathbf{\theta_{s}}-\mathbf{\psi}_{s}\right)\right\} -\frac{\rho}{2}\left\Vert \mathbf{\theta_{s}}-\mathbf{\psi}_{s}\right\Vert _{2}^{2}\nonumber \\
\mathbf{s.t.}\,\left|\mathbf{\theta}_{s,m}\right|\leq1\forall m\qquad\qquad\nonumber \\
\left|\mathbf{\psi}_{s,m}\right|=1\forall m\qquad\qquad\nonumber \\
\mathbf{\theta_{s}}^{H}\mathbf{B}_{ks}\mathbf{\theta_{s}}-\Gamma_{s,k_{s}}^{th}\theta_{s}^{H}\mathbf{A}_{ks}\mathbf{\theta_{s}}\geq0,\:\forall s,k_{s}\qquad\qquad
\end{eqnarray}

This formulation is more tractable and can be solved by updating $\mathbf{\theta_{s}},\mathbf{\psi}_{s},\mathbf{\zeta}_{s}$
, as follows:

 \emph{Update $\mathbf{\theta_{s}}$ }: the problem of optimizing
$\mathbf{\theta_{s}}$ with fixed $\mathbf{\psi}_{s},\mathbf{\zeta}_{s}$, can be expressed as
\begin{eqnarray}
\underset{\mathbf{\theta_{s}}}{\max}\,\stackrel[k_{s}=1]{K_{s}}{\sum}2\sqrt{\left(1+\alpha_{ks}\right)}\mathbf{Re}\left\{ \mathbf{\theta_{s}}^{H}\mathbf{\bar{B}}_{ks}\mathbf{y}_{m}\right\} \nonumber \\
-\stackrel[k_{s}=1]{K_{s}}{\sum}\mathbf{y}_{m}^{T}\left(\mathbf{\theta_{s}}^{H}\mathbf{C}_{ks}\mathbf{\theta_{s}}\right)\mathbf{y}_{m}\nonumber \\
-\mathbf{Re}\left\{ \mathbf{\zeta}_{s}^{H}\left(\mathbf{\theta_{s}}-\mathbf{\psi}_{s}\right)\right\} -\frac{\rho}{2}\left\Vert \mathbf{\theta_{s}}-\mathbf{\psi}_{s}\right\Vert _{2}^{2}\nonumber \\
\mathbf{s.t.}\,\left|\mathbf{\theta}_{s,m}\right|\leq1\forall m\qquad\qquad\nonumber \\
\mathbf{\theta_{s}}^{H}\mathbf{B}_{ks}\mathbf{\theta_{s}}-\Gamma_{s,k_{s}}^{th}\theta_{s}^{H}\mathbf{A}_{ks}\mathbf{\theta_{s}}\geq0,\:\forall s,k_{s}\qquad\qquad\label{eq:36}
\end{eqnarray}

This problem is convex and thus can be easily solved using software
tools as CVX. 

\emph{Update $\mathbf{\psi}_{s}$}: the problem of optimizing $\mathbf{\psi}_{s}$
with fixed $\mathbf{\theta_{s}},\mathbf{\zeta}_{s}$, can be written
as
\begin{eqnarray}
\underset{\mathbf{\psi}}{\max}\,-\stackrel[k_{s}=1]{K_{s}}{\sum}\mathbf{Re}\left\{ \mathbf{\zeta}_{s}^{H}\left(\mathbf{\theta_{s}}-\mathbf{\psi}_{s}\right)\right\} -\frac{\rho}{2}\left\Vert \mathbf{\theta_{s}}-\mathbf{\psi}_{s}\right\Vert _{2}^{2}\nonumber \\
\mathbf{s.t.}\,\left|\mathbf{\psi}_{s,m}\right|=1\forall m\qquad\qquad
\end{eqnarray}

which can be written as

\begin{eqnarray}
\underset{\mathbf{\psi}}{\max}\,-\stackrel[k_{s}=1]{K_{s}}{\sum}\mathbf{Re}\left\{ \left(\mathbf{\zeta}_{s}+\rho\mathbf{\theta_{s}}\right)^{H}\mathbf{\psi}_{s}\right\} -\frac{\rho}{2}\left\Vert \mathbf{\theta_{s}}\right\Vert _{2}^{2}\nonumber \\
-\frac{\rho}{2}\left\Vert \mathbf{\theta_{s}}\right\Vert _{2}^{2}-\mathbf{Re}\left\{ \mathbf{\zeta}_{s}\mathbf{\theta_{s}}\right\} \nonumber \\
\mathbf{s.t.}\,\left|\mathbf{\psi}_{s,m}\right|=1\forall m\qquad\qquad
\end{eqnarray}

\noindent where $\left\Vert \mathbf{\theta_{s}}\right\Vert _{2}^{2}=M$.
Thus, the problem can be further reduced to
\begin{eqnarray}
\underset{\mathbf{\psi}_{s}}{\max}\,\stackrel[k_{s}=1]{K_{s}}{\sum}\mathbf{Re}\left\{ \left(\mathbf{\zeta}_{s}+\rho\mathbf{\theta}_{ks}\right)^{H}\mathbf{\psi}_{ks}\right\} \nonumber \\
\mathbf{s.t.}\,\left|\mathbf{\psi}_{s,m}\right|=1\forall m\qquad\qquad
\end{eqnarray}

The optimal solution can be obtained as \cite{wideband1,wideband2,wideband3,wideband4}
\begin{equation}
\mathbf{\psi}_{s}^{*}=e^{j\angle\left(\mathbf{\zeta}_{s}+\rho\mathbf{\theta}_{s}\right)^{H}}.\label{eq:39}
\end{equation}
 \emph{Update $\mathbf{\zeta}_{s}$}: we can update the dual variable
$\mathbf{\zeta}_{s}$ by\cite{wideband1,wideband2,wideband3,wideband4}
\begin{equation}
\mathbf{\zeta}_{s}=\mathbf{\zeta}_{s}+\rho\left(\mathbf{\theta_{s}}-\mathbf{\psi}_{s}\right).\label{eq:40}
\end{equation}
Finally, the ADMM-based ideal RIS phase-shift design is
summarized in Algorithm 1. 

\begin{algorithm}
1.Initialize $\mathbf{\psi}_{s},\mathbf{\zeta}_{s}$ .

2. Repeat 

3. Calculate passive beamforming $\mathbf{\theta_{s}}$ from
(\ref{eq:36});

4. Calculate auxiliary variable $\mathbf{\psi}_{s}$ using (\ref{eq:39});

5. Update Lagrangian multiplier $\mathbf{\zeta}_{s}$ by (\ref{eq:40});

6. until convergence. 

\protect\caption{\label{alg:ADMM}ADMM based ideal RIS phase-shifts design algorithm.}

\end{algorithm}

\subsection{Service Selection Design }

Now with obtaining the ideal phase shifts, the optimization problem to obtain the service selection matrix A can be formulated as
\begin{eqnarray}
\underset{\mathbf{V}}{\max}\,\underset{s\in S}{\sum}\stackrel[k_{s}=1]{K_{s}}{\sum}\log_{2}\left(1+\frac{\left(N-K-1\right)P_{s,p}\mathbf{\theta_{s}}^{H}\mathbf{B}_{ks}\mathbf{\theta_{s}}}{\theta_{s}^{H}\mathbf{A}_{ks}\mathbf{\theta_{s}}\sigma_{k_{s}}^{2}}\right)\nonumber \\
\mathbf{s.t.}\,\mathbf{\theta}_{s}=e^{j\mathbf{\phi}_{s}\odot\mathbf{\tilde{v}}_{s}},\:\forall S\qquad\qquad\nonumber \\
\left\Vert \mathbf{v}_{m}\right\Vert _{0}\leq1,v_{s,m}\in\left\{ 0,1\right\} ,\:\forall s,m\nonumber \\
\frac{\mathbf{\theta_{s}}^{H}\mathbf{B}_{ks}\mathbf{\theta_{s}}}{\theta_{s}^{H}\mathbf{A}_{ks}\mathbf{\theta_{s}}}\geq\Gamma_{s,k_{s}}^{th},\:\forall s,k_{s}\qquad\qquad
\end{eqnarray}

which can be equivalently expressed as
\begin{eqnarray}
\underset{\mathbf{V}}{\max}\,\underset{s\in S}{\sum}\stackrel[k_{s}=1]{K_{s}}{\sum}\log_{2}\left(1+\frac{\mathbf{b}_{1}\mathbf{B}_{ks}\mathbf{b}_{1}^{T}}{\mathbf{b}_{1}\mathbf{A}_{ks}\mathbf{b}_{1}^{T}}\right)\qquad\qquad\qquad\nonumber \\
\mathbf{s.t.}\,\mathbf{\theta}_{s}=e^{j\mathbf{\phi}_{s}\odot\mathbf{\tilde{v}}_{s}},\:\forall S\qquad\qquad\nonumber \\
\left\Vert \mathbf{v}_{m}\right\Vert _{0}\leq1,v_{s,m}\in\left\{ 0,1\right\} ,\:\forall s,m\nonumber \\
\frac{\mathbf{b}_{1}\mathbf{B}_{ks}\mathbf{b}_{1}^{T}}{\mathbf{b}_{1}\mathbf{A}_{ks}\mathbf{b}_{1}^{T}}\geq\Gamma_{s,k_{s}}^{th},\:\forall s,k_{s}\qquad\qquad
\end{eqnarray}

\noindent where $\mathbf{b}_{1}=\left(\mathbf{\tilde{v}}_{s}^{H}\left(\textrm{diag}\left\{ e^{j\mathbf{\phi}_{s}^{H}}\right\} -\mathbf{I}_{M}\right)+\mathbf{1}\right)$.
\begin{IEEEproof}
\noindent The SINR can be reformulated as

\[
\frac{\mathbf{\theta_{s}}^{H}\mathbf{B}_{ks}\mathbf{\theta_{s}}}{\mathbf{\theta_{s}}^{H}\mathbf{A}_{ks}\mathbf{\theta_{s}}}=
\]

\[
\frac{\left(\mathbf{\tilde{v}}_{s}^{H}\odot e^{j\mathbf{\phi}_{s}^{H}}+\left(1-\mathbf{\tilde{v}}_{s}^{H}\right)\right)\mathbf{B}_{ks}\left(\mathbf{\tilde{v}}_{s}\odot e^{j\mathbf{\phi}_{s}}+\left(1-\mathbf{\tilde{v}}_{s}\right)\right)}{\left(\mathbf{\tilde{v}}_{s}^{H}\odot e^{j\mathbf{\phi}_{s}^{H}}+\left(1-\mathbf{\tilde{v}}_{s}^{H}\right)\right)\mathbf{A}_{ks}\left(\mathbf{\tilde{v}}_{s}\odot e^{j\mathbf{\phi}_{s}}+\left(1-\mathbf{\tilde{v}}_{s}\right)\right)}=
\]

\[
\frac{\left(\mathbf{\tilde{v}}_{s}^{H}\textrm{diag}\left\{ e^{j\mathbf{\phi}_{s}^{H}}\right\} +\left(1-\mathbf{\tilde{v}}_{s}^{H}\right)\right)\mathbf{B}_{ks}\left(\textrm{diag}\left\{ e^{j\mathbf{\phi}_{s}}\right\} \mathbf{\tilde{v}}_{s}+\left(1-\mathbf{\tilde{v}}_{s}\right)\right)}{\left(\mathbf{\tilde{v}}_{s}^{H}\textrm{diag}\left\{ e^{j\mathbf{\phi}_{s}^{H}}\right\} +\left(1-\mathbf{\tilde{v}}_{s}^{H}\right)\right)\mathbf{A}_{ks}\left(\textrm{diag}\left\{ e^{j\mathbf{\phi}_{s}}\right\} \mathbf{\tilde{v}}_{s}+\left(1-\mathbf{\tilde{v}}_{s}\right)\right)}=
\]

\[
\frac{\left(\mathbf{\tilde{v}}_{s}^{H}\left(\textrm{diag}\left\{ e^{j\mathbf{\phi}_{s}^{H}}\right\} -\mathbf{I}_{M}\right)+\mathbf{1}\right)\mathbf{B}_{ks}\left(\left(\textrm{diag}\left\{ e^{j\mathbf{\phi}_{s}}\right\} -\mathbf{I}_{M}\right)\mathbf{\tilde{v}}_{s}+\mathbf{1}\right)}{\left(\mathbf{\tilde{v}}_{s}^{H}\left(\textrm{diag}\left\{ e^{j\mathbf{\phi}_{s}^{H}}\right\} -\mathbf{I}_{M}\right)+\mathbf{1}\right)\mathbf{A}_{ks}\left(\left(\textrm{diag}\left\{ e^{j\mathbf{\phi}_{s}}\right\} -\mathbf{I}_{M}\right)\mathbf{\tilde{v}}_{s}+\mathbf{1}\right)}=
\]

\begin{equation}
\frac{\mathbf{b}_{1}\mathbf{B}_{ks}\mathbf{b}_{1}^{T}}{\mathbf{b}_{1}\mathbf{A}_{ks}\mathbf{b}_{1}^{T}}.
\end{equation}

\end{IEEEproof}
\textcolor{red}{} FP is applied to simplify the problem 
tractability. To apply FP method, we first define the dual variable
$\alpha=\left[\alpha_{1,1},...,\alpha_{S,K_{s}}\right]^{T}$ , thus
the problem can be written as
\begin{eqnarray}
\underset{\mathbf{V},\alpha}{\max}\, f_{1}\qquad\qquad\qquad\nonumber \\
\mathbf{s.t.}\,\mathbf{\theta}_{s}=e^{j\mathbf{\phi}_{s}\odot\mathbf{\tilde{v}}_{s}},\:\forall S\qquad\qquad\nonumber \\
\left\Vert \mathbf{v}_{m}\right\Vert _{0}\leq1,v_{s,m}\in\left\{ 0,1\right\} ,\:\forall s,m\nonumber \\
\frac{\mathbf{b}_{1}\mathbf{B}_{ks}\mathbf{b}_{1}^{T}}{\mathbf{b}_{1}\mathbf{A}_{ks}\mathbf{b}_{1}^{T}}\geq\Gamma_{s,k_{s}}^{th},\:\forall s,k_{s}
\end{eqnarray}

\noindent where $f_{1}=\underset{s\in S}{\sum}\stackrel[k_{s}=1]{K_{s}}{\sum}\log_{2}\left(1+\alpha_{ks}\right)-\underset{s\in S}{\sum}\stackrel[k_{s}=1]{K_{s}}{\sum}\alpha_{ks}+\underset{s\in S}{\sum}\stackrel[k_{s}=1]{K_{s}}{\sum}\frac{\left(1+\alpha_{ks}\right)\mathbf{b}_{1}\mathbf{B}_{ks}\mathbf{b}_{1}^{T}}{\mathbf{b}_{1}\mathbf{C}_{ks}\mathbf{b}_{1}^{T}}$
and $\mathbf{C}_{ks}=\mathbf{A}_{ks}+\mathbf{B}_{ks}$. Similarly,
with fixed $\mathbf{\theta_{s}}$, the optimal $\alpha_{ks}^{*}=\frac{\mathbf{b}_{1}\mathbf{B}_{ks}\mathbf{b}_{1}^{T}}{\mathbf{b}_{1}\mathbf{A}_{ks}\mathbf{b}_{1}^{T}}$.
For a given $\alpha_{ks}$ , by applying the quadratic transform in
\cite{FP1,FP2} the problem can be presented as
\begin{eqnarray}
\underset{\mathbf{V}}{\max}\, f_{2}\qquad\qquad\qquad\nonumber \\
\mathbf{s.t.}\,\mathbf{\theta}_{s}=e^{j\mathbf{\phi}_{s}\odot\mathbf{\tilde{v}}_{s}},\:\forall S\qquad\qquad\nonumber \\
\left\Vert \mathbf{v}_{m}\right\Vert _{0}\leq1,v_{s,m}\in\left\{ 0,1\right\} ,\:\forall s,m\nonumber \\
\frac{\mathbf{b}_{1}\mathbf{B}_{ks}\mathbf{b}_{1}^{T}}{\mathbf{b}_{1}\mathbf{A}_{ks}\mathbf{b}_{1}^{T}}\geq\Gamma_{s,k_{s}}^{th},\:\forall s,k_{s}
\end{eqnarray}

\noindent where $f_{2}=\underset{s\in S}{\sum}\stackrel[k_{s}=1]{K_{s}}{\sum}\frac{\left(1+\alpha_{ks}\right)\mathbf{b}_{1}\mathbf{B}_{ks}\mathbf{b}_{1}^{T}}{\mathbf{b}_{1}\mathbf{C}_{ks}\mathbf{b}_{1}^{T}}$,
which can be simplified to , $f_{3}=\underset{s\in S}{\sum}\stackrel[k_{s}=1]{K_{s}}{\sum}2\sqrt{\left(1+\alpha_{ks}\right)}\mathbf{Re}\left\{ \mathbf{y}_{m}^{T}\mathbf{\bar{B}}_{ks}\mathbf{b}_{1}^{T}\right\} -\stackrel[k_{s}=1]{K_{s}}{\sum}\mathbf{y}_{m}^{T}\mathbf{b}_{1}\mathbf{C}_{ks}\mathbf{b}_{1}^{T}\mathbf{y}_{m}$
\cite{FP1,FP2}. Thus, we can write the problem as
\begin{eqnarray}
\underset{\mathbf{V},\mathbf{y}}{\max}\, f_{3}\qquad\qquad\qquad\nonumber \\
\mathbf{s.t.}\,\mathbf{\theta}_{s}=e^{j\mathbf{\phi}_{s}\odot\mathbf{\tilde{v}}_{s}},\:\forall S\qquad\qquad\nonumber \\
\left\Vert \mathbf{v}_{m}\right\Vert _{0}\leq1,v_{s,m}\in\left\{ 0,1\right\} ,\:\forall s,m\nonumber \\
\mathbf{b}_{1}\mathbf{B}_{ks}\mathbf{b}_{1}^{T}\geq\Gamma_{s,k_{s}}^{th}\mathbf{b}_{1}\mathbf{A}_{ks}\mathbf{b}_{1}^{T}\:\forall s,k_{s}
\end{eqnarray}

With fixed $\mathbf{V}$ the optimal solution of $\mathbf{y}$ is
found to be \cite{FP1,FP2} $\mathbf{y}_{m}^{*}=\left(\mathbf{\theta_{s}}^{H}\mathbf{C}_{ks}\mathbf{\theta_{s}}\right)^{-1}\mathbf{\bar{B}}_{ks}\mathbf{\theta_{s}}$.
Each element of $\mathbf{V}$ is binary, $v_{s,m}\in\left\{ 0,1\right\} $,
the 0-norm constraint, $\left\Vert \mathbf{v}_{m}\right\Vert _{0}$,
can be equivalently transformed into, $\left\Vert \mathbf{v}_{m}\right\Vert _{0}=\left\Vert \mathbf{v}_{m}\right\Vert _{1}=\mathbf{1}^{T}\mathbf{v}_{m}\leq1$.
Then, invoking the difference-of-convex function method, the constraint
$v_{s,m}\in\left\{ 0,1\right\} $ can be transformed into $0\leq v_{s,m}\leq1$
and $\underset{s\in S}{\sum}\left(\mathbf{1}^{T}\mathbf{v}_{m}-\mathbf{v}_{m}^{T}\mathbf{v}_{m}\right)\leq0$.
We further apply the penalty and reformulate the problem as
\begin{eqnarray}
 & \underset{\mathbf{V}}{\max}\, f_{4}\nonumber \\
\mathbf{s.t.}\, & \mathbf{1}^{T}\mathbf{v}_{m}\leq1\forall m\nonumber \\
 & 0\leq v_{s,m}\leq1,\:\forall s,m\nonumber \\
 & \mathbf{b}_{1}\mathbf{B}_{ks}\mathbf{b}_{1}^{T}\geq\Gamma_{s,k_{s}}^{th}\mathbf{b}_{1}\mathbf{A}_{ks}\mathbf{b}_{1}^{T}\:\forall s\label{eq:51}
\end{eqnarray}

\noindent where $f_{4}=\underset{s\in S}{\sum}\stackrel[k_{s}=1]{K_{s}}{\sum}2\sqrt{\left(1+\alpha_{ks}\right)}\mathbf{Re}\left\{ \mathbf{y}_{m}^{T}\mathbf{\bar{B}}_{ks}\mathbf{b}_{1}^{T}\right\} -\stackrel[k_{s}=1]{K_{s}}{\sum}\mathbf{y}_{m}^{T}\left(\mathbf{b}_{1}\mathbf{C}_{ks}\mathbf{b}_{1}^{T}\right)\mathbf{y}_{m}-\tau\underset{s\in S}{\sum}\left(\mathbf{1}^{T}\mathbf{\tilde{v}}_{s}-\mathbf{\tilde{v}}_{s}^{T}\mathbf{\tilde{v}}_{s}\right)$
and $\tau>0$ is a penalty factor. By utilizing the first-order Taylor
expansion, the problem can be solved using software tools as CVX. 
\subsection{Sub-optimal Design}

In this sub-section, we provide some sub-optimal designs of an RIS
assisted multi-cell multi-band communication systems.

\subsubsection{QoS design}

Unlike the iterative solutions, here, we obtain the service selection
matrix $\mathbf{V}$ theoretically using the derived ergodic sum rate
expressions. Since the service selection matrix $\mathbf{V}$ is directly
affect the users QoS, we obtain \textbf{$\mathbf{V}$} that satisfies
QoS constraint of all users, i.e,. $\mathbf{b}_{1}\mathbf{B}_{ks}\mathbf{b}_{1}^{T}\geq\Gamma_{s,k_{s}}^{th}\mathbf{b}_{1}\mathbf{A}_{ks}\mathbf{b}_{1}^{T},\:\forall s,k_{s}$.
Thus, for user $k_{s}$ served by BS $S$, we can get $\frac{\mathbf{b}_{1}\mathbf{B}_{ks}\mathbf{b}_{1}^{T}}{\mathbf{b}_{1}\mathbf{A}_{ks}\mathbf{b}_{1}^{T}}=\Gamma_{s,k_{s}}^{th}$.
Let us express the two terms explicitly, the ratio simplifies to
\begin{equation}
\underset{i,j}{\sum}B_{i,j}b_{i}^{T}b_{j}-\Gamma_{s,k_{s}}^{th}\underset{i,j}{\sum}A_{i,j}b_{i}^{T}b_{j}=0\label{eq:55}
\end{equation}
\noindent where $A_{i,j}$, $B_{i,j}$ and $b_{j}$
are the elements of $\mathbf{A}_{ks}$, $\mathbf{B}_{ks}$ and $\mathbf{b}_{1}$,
respectively. We can find $\tilde{v}_{i}$ using simple linear search
algorithm, by activating the elements until the condition in (\ref{eq:55})
satisfied, e.g., search through all possible binary vector to find
the solution that satisfies (\ref{eq:55}).

\subsubsection{Time division (TD) protocol}
In this design, all the RIS elements serve only one BS in different
time slots. In each time slot, one BS transmits messages to the users
via the RIS, where the entire RIS is configured for the active BS.
Thus number of time slots, $T$ , is identical to number of BSs. The
achievable sum-rate for each BS in TD design can be evaluated by,
$R_{s}^{TD}=\frac{1}{S}R_{s}$. The optimal phase shifts of the RIS
at each time slot for each BS can be obtained by using the solution
in Section \ref{sub:Ideal-Phase-shift}.

\subsubsection{\noindent RIS elements division (ED) protocol}
\noindent In this design, the RIS elements are divided equally into
tiles, and each tile is assigned to a BS. Number of elements assigned
for each BS, the tile size, is $\frac{M}{S}$. The optimal phase shifts
of each tile can be obtained by using the solution in Section \ref{sub:Ideal-Phase-shift}.

\subsection{RSMA Power Allocation\label{sec:Rate-Maximization-through}}

The optimal power fraction for each BS, $t_{s}$, can be calculated
by solving the following maximization problem
\begin{equation}
\underset{0\leq t_{s}\leq1}{\max}\quad\left(\min_{j}\left(\mathcal{E}\left\{ R_{s,j}^{c}\right\} \right)_{j=1}^{K_{s}}+\stackrel[k=1]{K_{s}}{\sum}\mathcal{E}\left\{ R_{s,k}^{p}\right\} \right)\label{eq:73}
\end{equation}

The optimal value of $t_{s}$ can be determined using the first-order (derivative) condition. However, given the ergodic rate expressions, the solution
would be complicated. A basic one-dimensional search methods such
as golden section (GS) technique over the period $0\leq t_{s}\leq1$ can
be applied to obtain the optimal power fraction $t_{s}$. GS 
algorithm to reach the optimal $t_{s}$ is presented in Algorithm
\ref{alg:GS}. 
\begin{algorithm}
1: Initialize $\varrho=0,\,\lambda=1,\textrm{ and }\zeta=\frac{-1+\sqrt{5}}{2}.$

2: Repeat

3: Update $t_{s1}=\varrho+(1-\zeta)\lambda\textrm{ and }t_{s2}=\lambda+(1-\zeta)\varrho.$

4: Obtain $R\left(t_{s1}\right)$ and $R\left(t_{s2}\right)$ from (\ref{eq:5-1}).

5:If $R\left(t_{s1}\right)>R\left(t_{s2}\right)$, set $\varrho=t_{s1}$.
Else set $\zeta=t_{s2}$.

6:Until $\left|\varrho-\lambda\right|$ converges.

7: Find $ts^{*}=(\varrho+\lambda)/2.$

\protect\caption{GS method\label{alg:GS}.}
\end{algorithm}

To reduce the complexity, an effective
power allocation scheme has been proposed in \cite{pars6}. Basically,
 a part ($t_{s}$) of the total power is allocated to the private
signals to perform approximately the same sum rate as that achieved
by the conventional multiple-user transmission, NoRS, with full power.
Then, the rest of the total power can be allocated to the common message
\cite{pars6}. The sum-rate gain of RSMA over NoRS can be evaluated
by $\varDelta R_{s}=\mathcal{E}\left\{ R_{s,c}\right\} +\stackrel[k=1]{K_{s}}{\sum}\left(\mathcal{E}\left\{ R_{s,k}^{p}\right\} -\mathcal{E}\left\{ R_{k}^{NoRS}\right\} \right)$.
Thus, the optimal ratio, $t_{s}$, can be obtained by achieving the
equality, $\mathcal{E}\left\{ R_{s,k}^{p}\right\} \approx\mathcal{E}\left\{ R_{k}^{NoRS}\right\} $. 

Finally, according to the above analysis, the total steps to obtain the optimal practical RIS design and RSMA power allocation are summarized in Algorithm 3.

\begin{algorithm}
1: Initialize ,$\mathbf{V}$, $\mathbf{\psi}_{s},\mathbf{\zeta}_{s},\alpha,\mathbf{y}$
and $\theta_{s}$;

2: for $s=1$ to $S$ do; 

4: Calculate dual and auxiliary variables;

5: Update ideal phase-shifts by Algorithm \ref{alg:ADMM};

6: end for;

6: Obtain \textbf{$\mathbf{V}$} by solving (\ref{eq:51});

7: Repeat until convergence.

8: Allocate the power by Algorithm \ref{alg:GS}.

9:Return , $\mathbf{\phi}$,$\mathbf{V}$, and $t_{s}$.

\protect\caption{Practical RIS Design and RSMA power allocation for sum-rate maximization
problem.}

\end{algorithm}

\section{Numerical Results\label{sec:Numerical-Results}}

\begin{figure}[!t]
\noindent \begin{centering}
\includegraphics[bb=320bp 110bp 650bp 470bp,clip,scale=0.25]{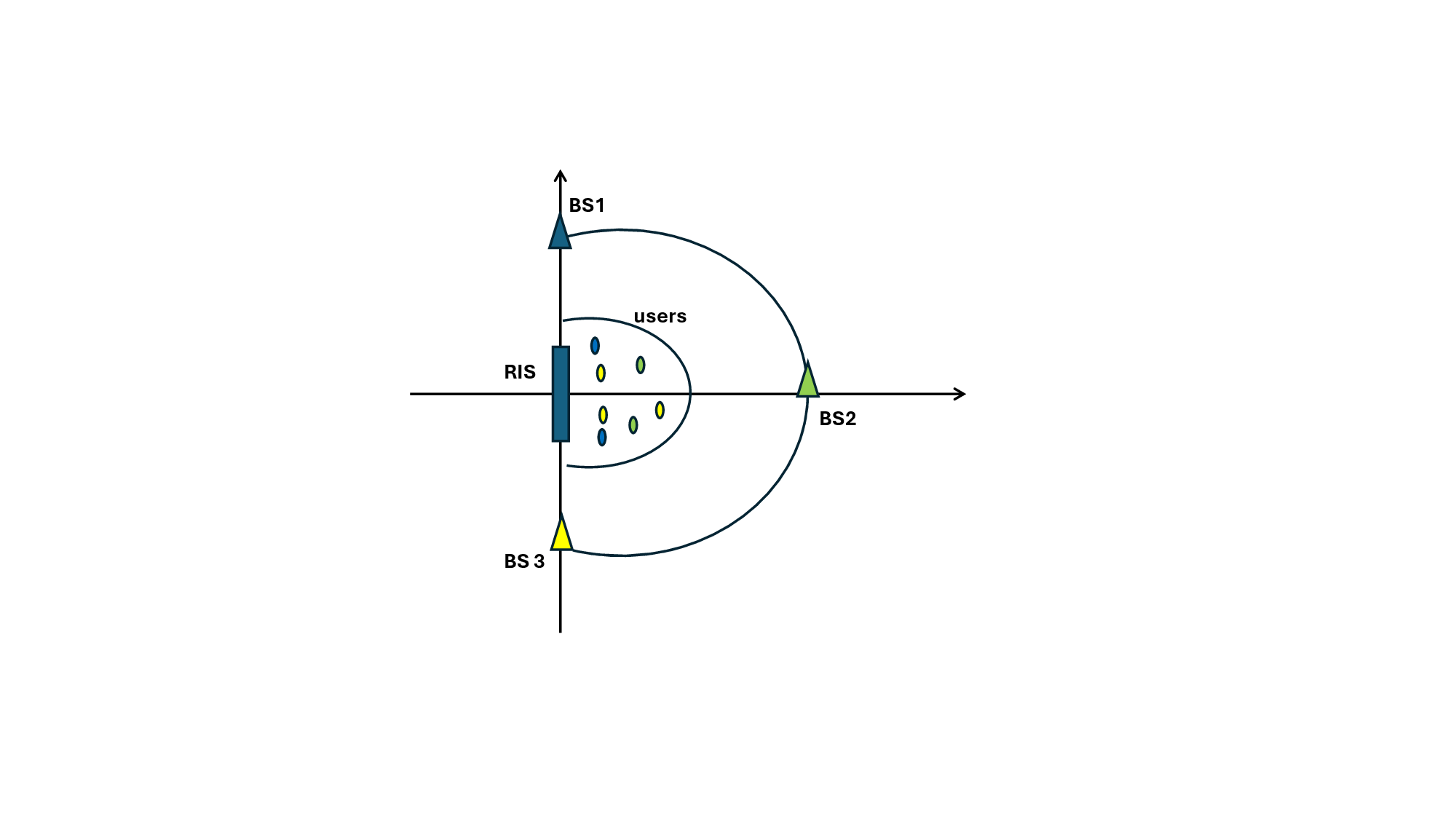}
\par\end{centering}

\protect\caption{\label{fig:numRes}An illustration of the RIS, BSs, and users locations.}
\end{figure}

In this section, simulation and numerical results are presented to
confirm the accuracy of the derived analytical expressions and demonstrate
the effectiveness of the proposed algorithms. It is assumed that the
multi-cell system consists of $S=3$ BSs/cells, and the users experience similar noise variance, $\sigma^{2}$, 
thus the transmit SNR at each BS is $\gamma=\frac{P_{s}}{\sigma^{2}}$.
The QoS requirement of each user is $\Gamma_{s,k_{s}}^{th}=5\textrm{ dB}\:\forall s,k_{s}$.
A two-dimensional coordinate system is considered where the RIS is
situated at (0, 0), and the BSs are located $50\textrm{ m}$
away from the RIS. In addition, the users in each cell are located
$2\textrm{ m}$ away from the RIS as shown in Fig. \ref{fig:numRes}.

\subsection{Ergodic sum-rate analysis }

\begin{figure}
\noindent \begin{centering}
\subfloat[\label{fig:1a}Sum-rate versus transmit SNR for different number of
BSs antennas when $M=5$.]{\noindent \begin{centering}
\includegraphics[scale=0.28]{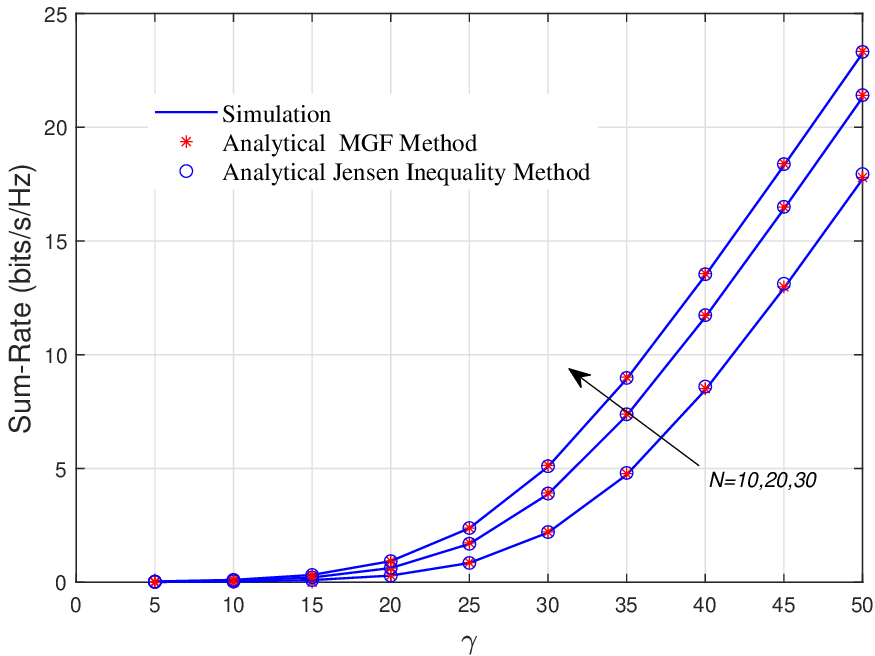}
\par\end{centering}

}
\par\end{centering}

\noindent \begin{centering}
\subfloat[\label{fig:1b}Sum-rate versus transmit SNR for different number of
RIS elements when $N_{s}=30$.]{\noindent \begin{centering}
\includegraphics[scale=0.28]{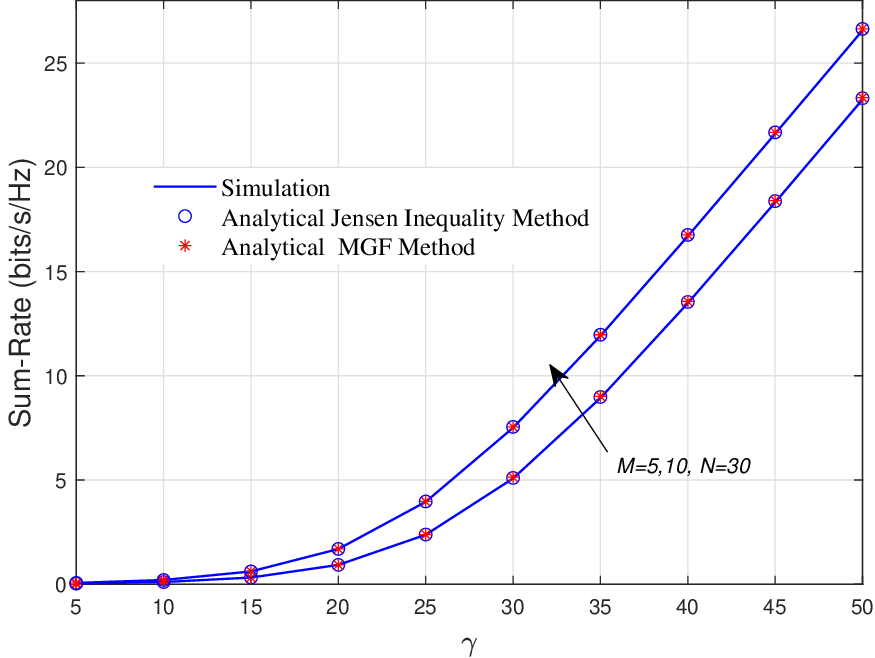}
\par\end{centering}

}
\par\end{centering}

\protect\caption{\label{fig:1}Sum-rate versus transmit SNR for different numbers of
BS antennas and RIS elements. }
\end{figure}

Firstly, to validate the ergodic sum-rates expressions derived in
this work based on MGF and Jensen inequality methods, in Fig. \ref{fig:1},
we plot the ergodic sum-rate with respect to the transmit SNR, $\gamma$.
Fig. \ref{fig:1a}, presents the ergodic sum-rate for different numbers
of BS antennas, $N_{s}=10,20\textrm{ and }30$, and number of RIS
elements is $M=5$, while Fig. \ref{fig:1b}, shows the ergodic sum-rate
with when $N_{s}=30,\,\textrm{ and }M=5\textrm{ and }10$. In these
figures, the simulation results are presented by solid lines, while
the analytical results based on MGF and Jensen inequality approaches
are presented by star and circle markers, respectively. It is evident that the simulation and analytical results are in good agreement, thereby validating the derivations presented
 in Section (\ref{sec:Ergodic-Rates-Analysis}).
It is also apparent that the achievable sum-rate enhances with increasing
the transmit SNR. From Figs. \ref{fig:1a} and \ref{fig:1b}, it can
be noted that using large number of BS antennas $N_{s}$ and RIS elements
$M$ leads to improve the achievable sum-rate.

\begin{figure}
\noindent \begin{centering}
\includegraphics[scale=0.28]{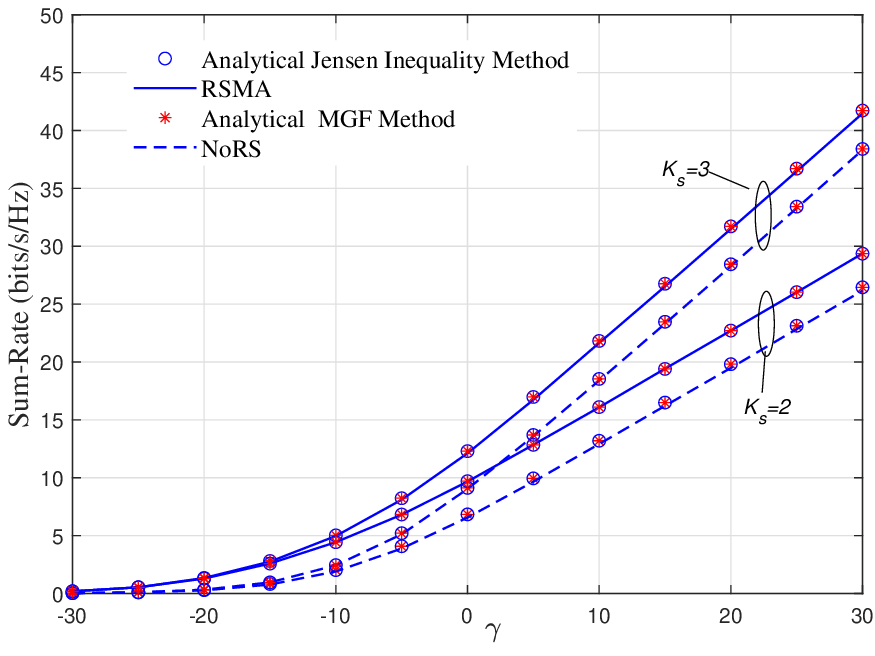}
\par\end{centering}

\protect\caption{\label{fig:2}Sum-rate versus transmit SNR with different numbers
of users $K_{s}$ for RSMA and NoRS schemes.}

\end{figure}

Furthermore, in Fig. \ref{fig:2}, we illustrate the ergodic sum-rate
of the RSMA and NoRS schemes versus transmit SNR for different numbers
of users, $K_{s}=2\textrm{ and }3$, when $N_{s}=15$, $M=5$ and
the distances are normalized. The results in this figure explain clearly
the benefits of using RSMA over the conventional NoRS schemes, and
justify the implementation of RSMA in such RIS communication systems.
RSMA scheme outperforms the conventional NoRS technique with almost
an up to 5dB gain in the SNR for a given sum rate. In addition, as
expected increasing number of users always enhances the achievable
sum-rates for both schemes. 

\subsection{Ideal RIS design }

\begin{figure}
\noindent \begin{centering}
\includegraphics[scale=0.28]{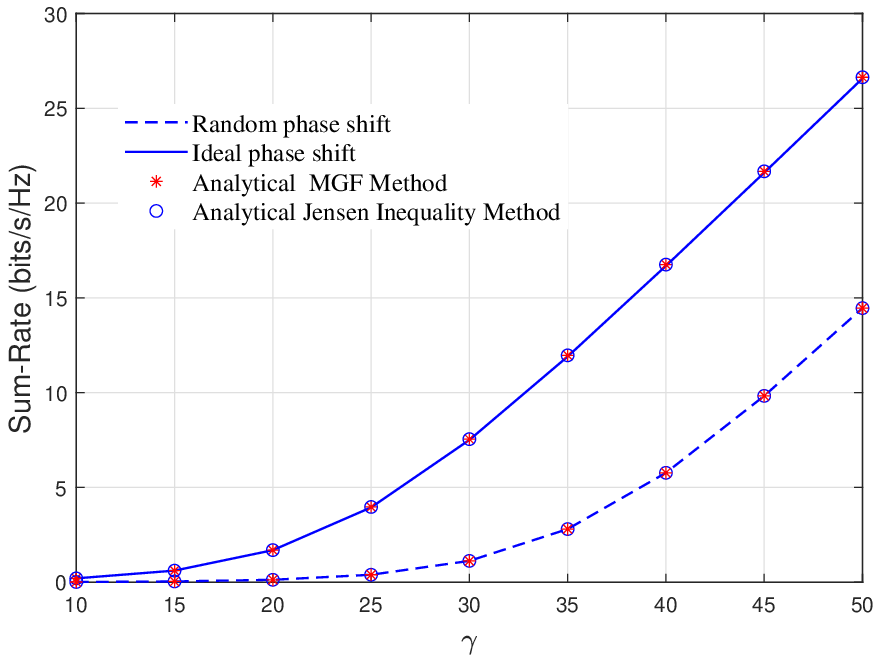}
\par\end{centering}

\protect\caption{\label{fig:3}Ergodic sum-rates versus transmit SNR,$\gamma$, for
different phase shifts.}

\end{figure}

Fig. \ref{fig:3} plots the achievable sum-rates with both optimal and random phase shifts when $N_{s}=30$ and $M=10$. It can be observed
that the ergodic sum-rate utilizing the optimal phase shifts is much
higher than the ergodic sum-rate utilizing random phase shifts. Thus, the optimal phase shifts result in significant
SNR gains compared to the random phase shifts. The gain attained using the optimal phase shifts over the random phase shifts is about 12dB for a given sum-rate value.

\begin{figure}
\noindent \begin{centering}
\subfloat[\label{fig:4a}Sum-rate vs. transmit SNR with low Qos requirements.]{\noindent \begin{centering}
\includegraphics[scale=0.28]{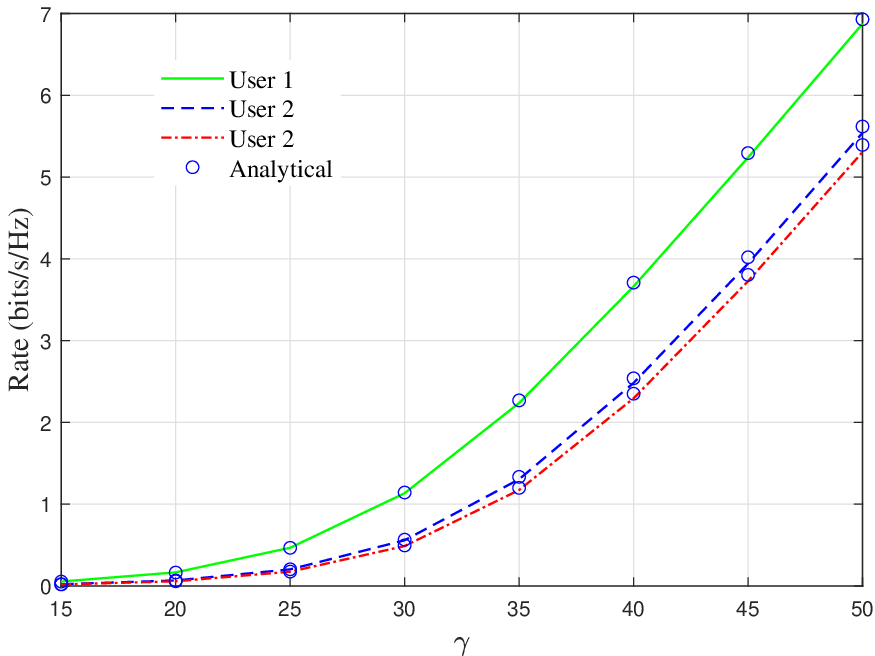}
\par\end{centering}

}
\par\end{centering}

\noindent \begin{centering}
\subfloat[\label{fig:4b}Sum-rate vs. transmit SNR with high QoS requirement
at user 3.]{\noindent \begin{centering}
\includegraphics[scale=0.28]{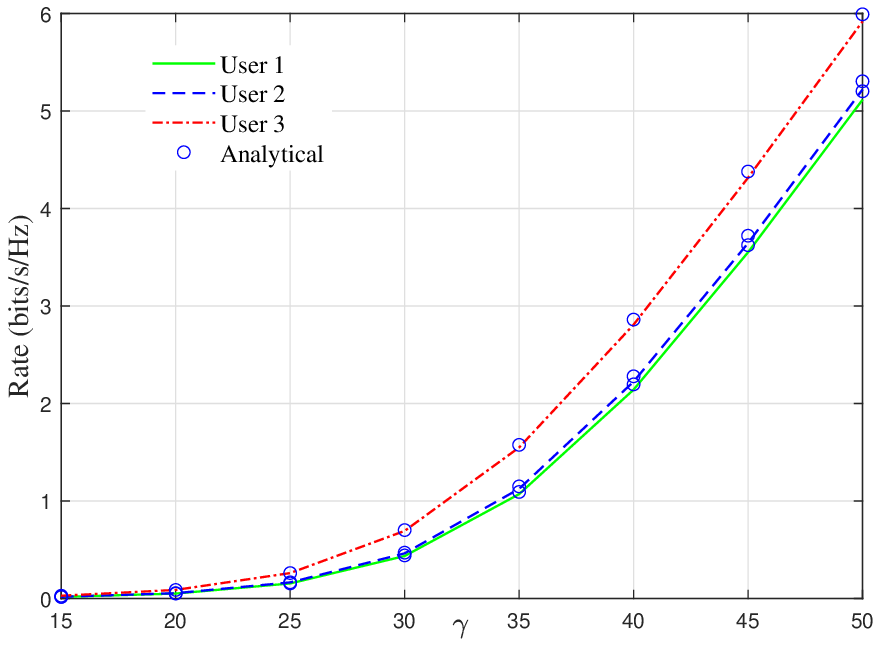}
\par\end{centering}

}
\par\end{centering}

\protect\caption{\label{fig:4}Sum-rate vs. transmit SNR for different QoS requirements.}

\end{figure}

In addition, Fig. \ref{fig:4} illustrates the achievable users rates
 for different QoS requirements, $\Gamma_{s,k_{s}}$,
when $N_{s}=10$, and $M=5$. Firstly, in Fig. \ref{fig:4a} we present
the users rates when the QoS requirement of each user is relatively
low and achievable. In this case, to maximize the sum-rate, the RIS
phase shifts have been designed to reflect most of the beam toward
the best user's channel which is in this case user 1, while the other
two users achieve almost similar rates. Then, in Fig. \ref{fig:4b}
we increase the QoS requirement of the weak user, user 3. As we can
see, in this case, the RIS is designed to provide more power toward
user 3 and thus enhance the achievable rate at user 3 on the expenses
of user 1 performance and the achievable sum-rate. It is evident from
these results that the QoS constraints should be adjusted to provide
fairness between the users.

\subsection{Practical RIS design }

\begin{figure}
\noindent \begin{centering}
\subfloat[\label{fig:5a}Total sum-rate versus transmit SNR when $S=2$.]{\noindent \begin{centering}
\includegraphics[scale=0.28]{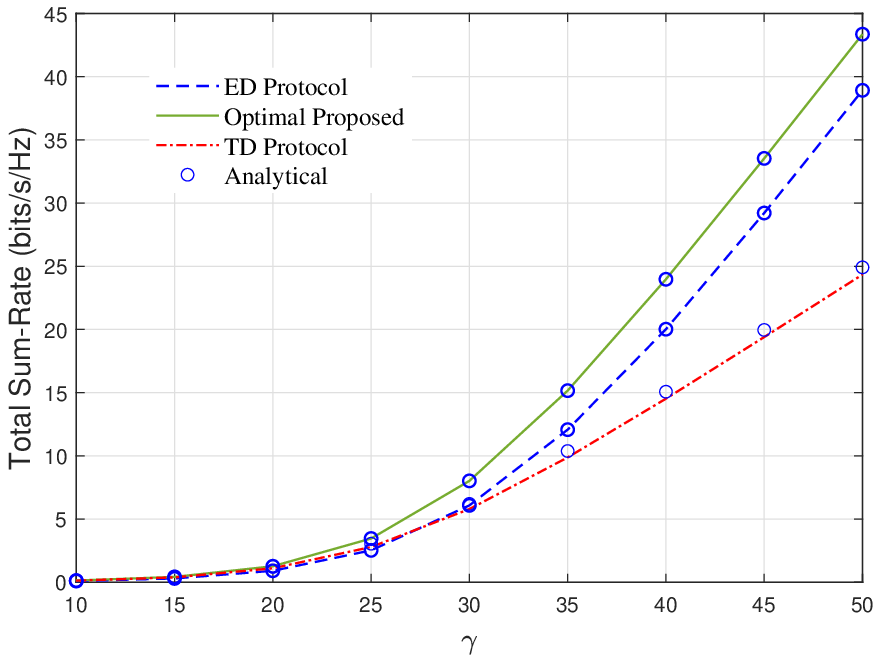}
\par\end{centering}

}
\par\end{centering}

\noindent \begin{centering}
\subfloat[\label{fig:5b}Total sum-rate vs. transmit SNR when $S=3$.]{\noindent \begin{centering}
\includegraphics[scale=0.28]{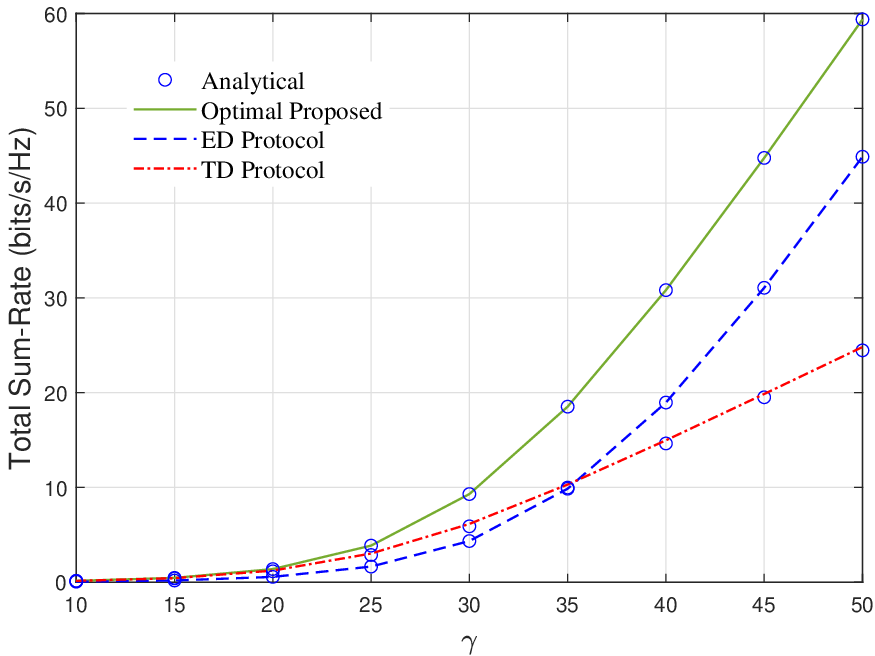}
\par\end{centering}

}
\par\end{centering}

\protect\caption{\label{fig:5}Total sum-rate vs. transmit SNR for different number
of BSs.}

\end{figure}

Fig. \ref{fig:5} shows the total sum-rates versus the transmit SNR
for two scenarios $S=2$ and $S=3$ when $M=10,N_{s}=30,\textrm{ and }K=3$.
Fig. \ref{fig:5a} shows the case when $S=2$ and Fig. \ref{fig:5b}
shows the case $S=3$. Additionally, to provide fair comparisons,
results of ED and TD protocols are added as benchmarks. It can be
seen that the proposed algorithm has greater performance than
the other schemes under different settings. The proposed algorithm
 provides up to 3dB gain when $S=2$ and 5dB gain when $S=3$ compared
with the ED protocol and 10 dB gain when $S=2$ and 15 dB gain when
$S=3$ compared with the TD protocol.

\begin{figure}
\noindent \begin{centering}
\subfloat[\label{fig:6a}Total sum-rate versus the BS-RIS distance.]{\noindent \begin{centering}
\includegraphics[scale=0.28]{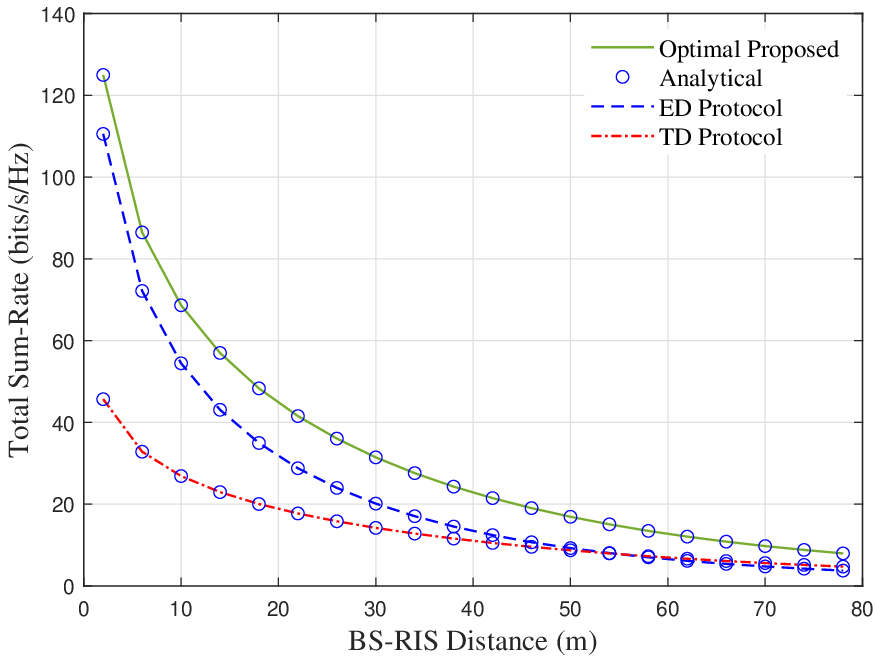}
\par\end{centering}

}
\par\end{centering}

\noindent \begin{centering}
\subfloat[\label{fig:6b}Total sum-rate versus the RIS-users distance.]{\noindent \begin{centering}
\includegraphics[scale=0.28]{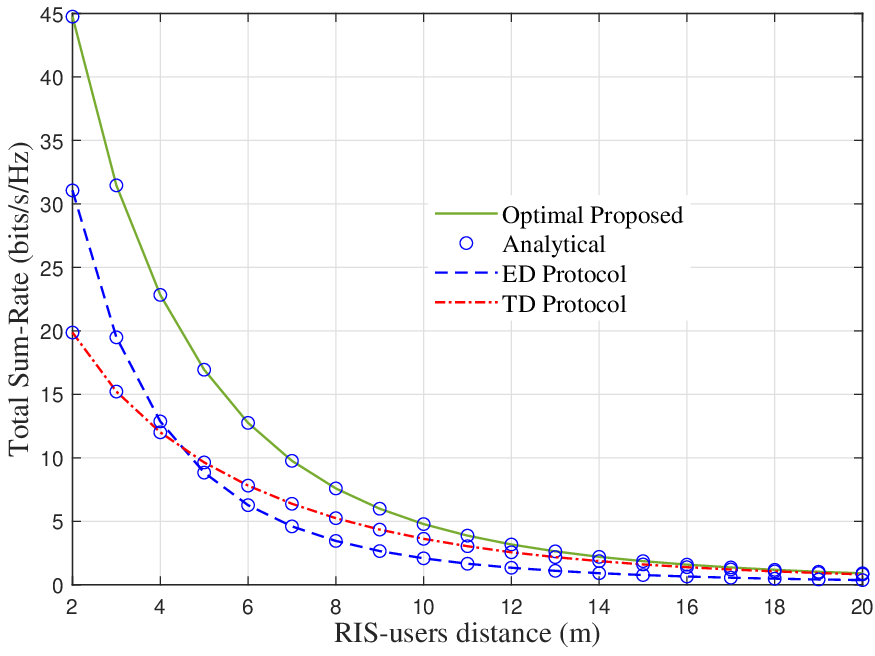}
\par\end{centering}

}
\par\end{centering}

\protect\caption{\label{fig:6}Total sum-rate vs. the BS-RIS distance, and RIS-users
distance when $\gamma=45\textrm{dB}$.}

\end{figure}

To consider the effect of the RIS location on the system performance,
we illustrate the total sum-rate versus the BS-RIS distance in Fig. \ref{fig:6} (a) and versus the RIS-user
distance in Fig. \ref{fig:6} (b). It can be
observed that when the RIS is located far away from the BS and/or 
the users, the achievable sum-rates deteriorate. This is because the reflected signals from the RIS become weaker due 
to larger path-loss. Having said that, the sum-rate is more sensitive to the RIS-user distance than BS-RIS distance, as we can notice the sum-rate degrades fast as the users move away from the RIS. Increasing the transmit power, number of BS antennas and/or  RIS elements can tackle this issue. Our proposed algorithm always performs better than the other schemes under different RIS and users' locations.

\begin{figure}
\noindent \begin{centering}
\includegraphics[scale=0.28]{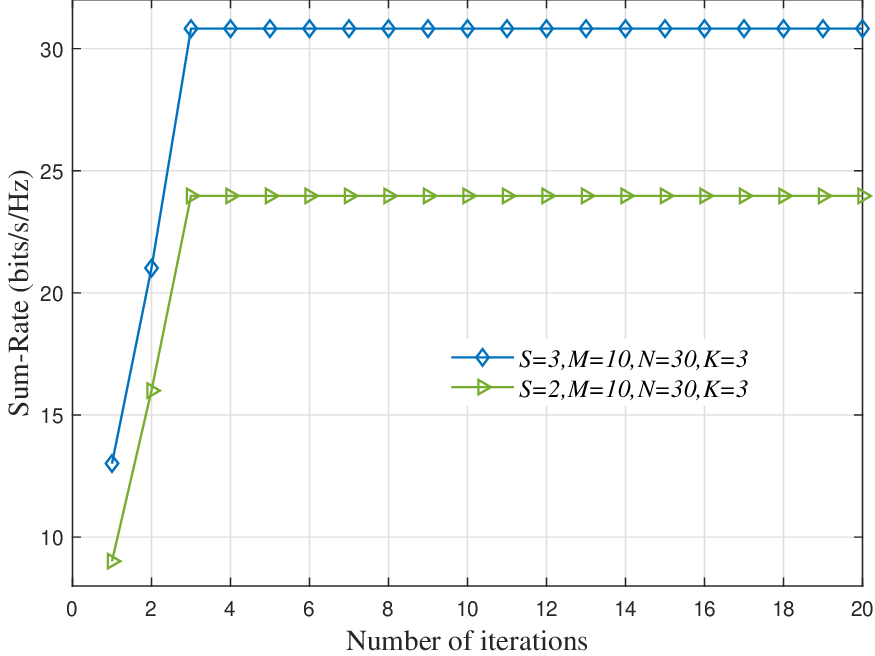}
\par\end{centering}

\protect\caption{\label{fig:7}Convergence of the proposed algorithm.}
\end{figure}

To illustrate the convergence of the proposed algorithm, the total sum-rate versus the number of iterations is plotted in Fig. \ref{fig:7}
when $N_{s}=30$, $M=10$, $\gamma=40\textrm{dB}$ for $S=2$ and
$3$. As we can see, the convergence is reached after
a limited number of iterations under different system parameters.

\subsection{RSMA Power Allocation }

\begin{figure}[!t]
\noindent \begin{centering}
\includegraphics[scale=0.28]{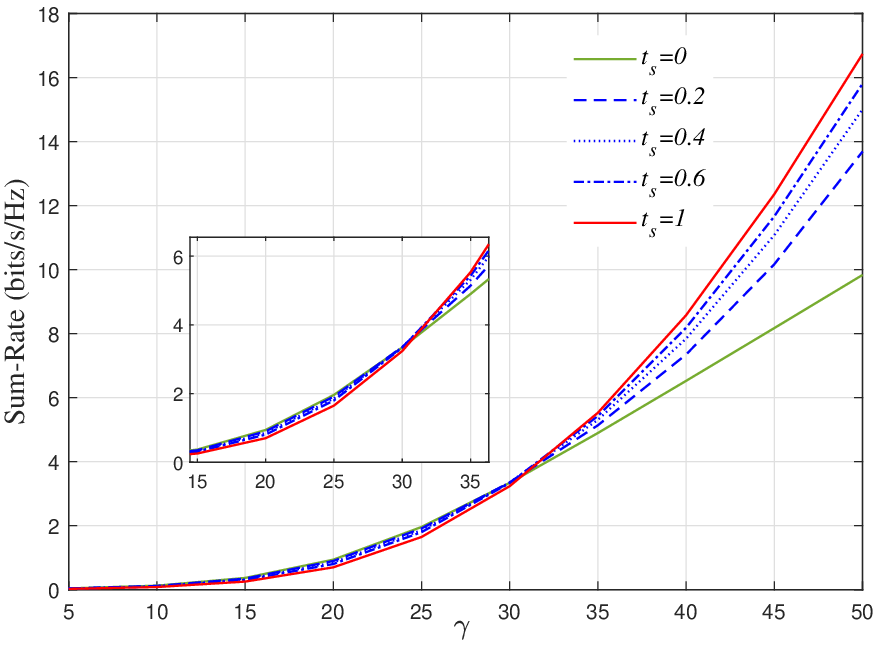}
\par\end{centering}

\protect\caption{\label{fig:8}Sum-rate vs. transmit SNR for various values of $t_{s}$.}
\end{figure}

In Fig. \ref{fig:8}, we plot the achievable sum-rate
versus transmit SNR for different values of $t_{s}$, when $N_{s}=30,\, M=10,$
and $K=3$. Interestingly enough, it can be observed that at low and
medium SNR values, $\textrm{SNR}\leq32\textrm{ dB}$, the optimal value
of $t_{s}$ in these ranges is approximately close to 0. This means,
in this case, most of the power should be allocated to the common
part. However, at high SNR values, $\textrm{SNR}\geq32\textrm{ dB}$,
the achievable sum-rate enhances as the power fraction $t_{s}$ increases.Thus, in this SNR range, most of the power should be allocated to the private part.

\section{Conclusion\label{sec:Conclusion} }
In this work, we first provided two methods to derive the ergodic sum-rate for RIS-assisted multi-cell multi-band RSMA communication systems. Then based on the analytical derivation, we investigated the RIS reflection design and RSMA power allocation scheme for the total sum-rate maximization problem. Efficient algorithms have been proposed to solve the total sum-rate optimization problem by exploiting BCD, ADMM, and line search algorithms. Numerical results showed that significant performance improvement can be attained using the proposed algorithms, which converge after a few iterations. Additionally, the users' performance can be controlled by adjusting the phase shifts according to their QoS requirements. 


\section*{Appendix A}

\noindent For any $u,v>0$, the ergodic rate can be calculated using the MGFs by \cite{usefullemma}  
\begin{equation}
\mathcal{E}\left[\textrm{ln}\left(1+\frac{u}{v}\right)\right]=\overset{\infty}{\underset{0}{\int}}\frac{1}{z}\left(\mathscr{\mathcal{M}}_{v}\left(z\right)-\mathscr{\mathcal{M}}_{u,v}\left(z\right)\right)dz,\label{eq:usefulemma-1}
\end{equation}
where $\mathscr{\mathcal{M}}_{u}\left(z\right)=\mathcal{E}\left[e^{-zu}\right]$
is the MGF of $u$ and $\mathscr{\mathcal{M}}_{u,v}\left(z\right)=\mathcal{E}\left[e^{-z\left(u+v\right)}\right]$.
Accordingly, (\ref{eq:13}) can be expressed as $\gamma_{s,k_{s}}^{c}=\frac{u}{\upsilon+\sigma_{k_{s}}^{2}}$
where $u=P_{s,c}\, x$, $x=\mathbf{g}_{k_{s}}\mathbf{g}_{k_{s}}^{H}$,
$\upsilon=P_{s,p}y$ and $y=\frac{1}{\left[\left(\mathbf{G}_{s}\mathbf{G}_{s}^{H}\right)^{-1}\right]_{k,k}}$.
Using (\ref{eq:usefulemma-1}), the ergodic rate of the common
message at the $k_{s}$ user is
\begin{equation}
\bar{R}_{s,c_{k_{s}}}=\frac{1}{\ln\left(2\right)}\,\overset{\infty}{\underset{0}{\int}}\frac{1}{z}\left(1-\mathscr{\mathcal{M}}_{u}\left(z\right)\right)\mathscr{\mathcal{M}}_{\upsilon}\left(z\right)e^{-z\sigma_{k_{s}}^{2}}\, dz.\label{eq:65}
\end{equation}
Since $\mathbf{g}_{k_{s}}\sim\mathcal{CN}\left(\sqrt{\frac{\kappa_{s}}{1+\kappa_{s}}}\mathbf{h}_{r,k_{s}}^{H}\Theta_{s}\bar{\mathbf{H}}_{s},\frac{\kappa_{s}}{1+\kappa_{s}}\mathbf{h}_{r,k_{s}}^{H}\mathbf{h}_{r,k_{s}}^{H}\right)$,
thus $x=\mathbf{g}_{k_{s}}\mathbf{g}_{k_{s}}^{H}$ has Gamma distribution
with probability density function (PDF) given by $f\left(x\right)=\frac{\left(2\nu\right)^{\frac{-N}{2}}}{\Gamma\left(\frac{N}{2}\right)}x^{\frac{N}{2}-1}e^{\left(\frac{-x}{2\nu}\right)}$,
where $\nu=\frac{1}{1+\kappa_{s}}\mathbf{h}_{r,k_{s}}^{H}\mathbf{h}_{r,k_{s}}+\frac{\kappa_{s}}{1+\kappa_{s}}\mathbf{h}_{r,k_{s}}^{H}\Theta_{s}\bar{\mathbf{H}}_{s}^{H}\bar{\mathbf{H}}_{s}\Theta_{s}^{H}\mathbf{h}_{r,k_{s}}$.
Thus, the MGF of $x$ is $\mathscr{\mathcal{M}}_{x}\left(z\right)=\mathbb{E}\left[e^{-zx}\right]=\left(1+2\nu z\right)^{\frac{-N}{2}}$.
We also know that the PDF of $\zeta=\left[\left(\mathbf{G}_{s}\mathbf{G}_{s}^{H}\right)^{-1}\right]_{k,k}$
is $f_{\zeta}\left(\zeta\right)=\frac{\left(\Psi_{k}\right)^{N-K+1}\, e^{\frac{-\Psi_{k}}{\zeta}}}{\Gamma\left(N-K+1\right)\,\zeta^{\left(N-K+2\right)}}$,
where $\Psi_{k}$ is the $k\textrm{th }$ diagonal element of $\Psi_{k}=\left[\left(\left(\frac{1}{1+\kappa_{s}}\right)\mathbf{H}_{r}^{H}\mathbf{H}_{r}+\left(\frac{\kappa_{s}}{1+\kappa_{s}}\right)\mathbf{H}_{r}^{H}\Theta_{s}\mathbf{a}_{M}^{H}\mathbf{a}_{M}\Theta_{s}^{H}\mathbf{H}_{r}\right)^{-1}\right]_{k_{s},k_{s}}$.
Let $y=\frac{1}{\zeta}$, thus the PDF of $y$ is $f_{Y}\left(y\right)=\frac{\left(\Psi_{k}\right)^{N-K+1}\, y^{\left(N-K\right)}e^{-\Psi_{k}y}}{\Gamma\left(N-K+1\right)}$, thus
the MGF of $y$ is $\mathcal{M}_{Y}\left(z\right)=\left(\Psi_{k}\right)^{N-K+1}\left(\Psi_{k}+z\right)^{K-N-1}$.
By substituting the values of $\mathscr{\mathcal{M}}_{x}\left(z\right)$
and $\mathscr{\mathcal{M}}_{y}\left(z\right)$ into (\ref{eq:65})
we can get

\[
\bar{R}_{s,c_{k_{s}}}=\frac{1}{\ln\left(2\right)}\,\overset{\infty}{\underset{0}{\int}}\frac{1}{z}\left(1-\left(1+2\nu P_{s,c}z\right)^{\frac{-N}{2}}\right)
\]

\begin{equation}
\times\left(\Psi_{k}\right)^{N-K+1}\left(\Psi_{k}+P_{s,p}z\right)^{K-N-1}e^{-z\beta}\, dz.
\end{equation}

By Applying Gaussian rules the ergodic common rate in Lemma~1 can
be obtained. Now, the ergodic private-rate is 
\[
\mathcal{E}\left[R_{k}^{p}\right]=\overset{\infty}{\underset{0}{\int}}\log_{2}\left(1+\frac{P_{s,p}y}{\sigma_{k_{s}}^{2}}\right)
\]

\begin{equation}
\times\frac{y^{\left(N_s-K\right)}\left(\Psi_{k}\right)^{N_s-K+1}\, e^{-\Psi_{k}y}}{\Gamma\left(N_s-K+1\right)\,}dy.\label{eq:31}
\end{equation}

By applying Gaussian rules the ergodic private rate in Lemma~2 can
be obtained.

\section*{Appendix B}

\textcolor{blue}{} The average SINR of the common message at
the $k_{s}^{th}$ user is 

\begin{equation}
\mathcal{E}\left\{ \gamma_{s,k_{s}}^{c}\right\} =\frac{P_{s,c}\,\mathcal{E}\left\{ x\right\} }{P_{s,p}\mathcal{E}\left\{ y\right\} +\sigma_{k_{s}}^{2}},\label{eq:14}
\end{equation}

\noindent where $x=\eta_{k_{s}}\left\Vert \mathbf{g}_{k_{s}}\right\Vert ^{4}$,
$\eta_{k_{s}}=\frac{1}{K_{s}\left(\frac{\kappa_{s}}{\kappa_{s}+1}\mathbf{\bar{h}}_{r,k_{s}}\Theta_{s}\mathbf{\bar{H}}_{s}\mathbf{\bar{H}}_{s}^{H}\Phi_{s}^{H}\mathbf{\bar{h}}_{r,k_{s}}+\frac{N}{\kappa_{s}+1}M\right)}$
and $y=\frac{1}{\left[\left(\mathbf{G}_{s}\mathbf{G}_{s}^{H}\right)^{-1}\right]_{k_{s},k_{s}}}$.
Now, the average of $x$ can be evaluated by, 

\begin{equation}
\mathcal{E}\left\{ x\right\} =\eta_{k_{s}}\mathcal{E}\left\Vert \mathbf{g}_{k_{s}}\right\Vert ^{4}=\eta_{k_{s}}\mathcal{E}\left\{ \left|\mathbf{h}_{r,k_{s}}^{H}\Theta_{s}\mathbf{H}_{s}\mathbf{H}_{s}^{H}\Theta_{s}^{H}\mathbf{h}_{r,k_{s}}\right|^{2}\right\} 
\end{equation}

\noindent where
\[
\mathbf{h}_{r,k_{s}}^{H}\Theta_{s}\mathbf{H}_{s}\mathbf{H}_{s}^{H}\Theta_{s}^{H}\mathbf{h}_{r,k_{s}}=\mathbf{h}_{r,k}^{H}\Theta_{s}\left(\frac{\kappa_{s}}{\kappa_{s}+1}\mathbf{\bar{H}}\mathbf{\bar{H}}^{H}+\right.
\]
\[
\left.\frac{\sqrt{\kappa_{s}}}{\kappa_{s}+1}\mathbf{\bar{H}}\mathbf{\tilde{H}}^{H}+\frac{\sqrt{\kappa_{s}}}{\kappa_{s}+1}\mathbf{\tilde{H}}\mathbf{\bar{H}}^{H}+\frac{1}{\kappa_{s}+1}\mathbf{\tilde{H}}\mathbf{\tilde{H}}^{H}\right)\Theta_{s}^{H}\mathbf{h}_{r,k_{s}}
\]
\[
=\frac{1}{\kappa_{s}+1}\mathbf{h}_{r,k}^{H}\Theta_{s}\left(\kappa_{s}\mathbf{\bar{H}}\mathbf{\bar{H}}^{H}+\sqrt{\kappa_{s}}\mathbf{\bar{H}}\mathbf{\tilde{H}}^{H}\right.
\]

\begin{equation}
\left.+\sqrt{\kappa_{s}}\mathbf{\tilde{H}}\mathbf{\bar{H}}^{H}+\mathbf{\tilde{H}}\mathbf{\tilde{H}}^{H}\right)\Theta_{s}^{H}\mathbf{h}_{r,k_{s}}=\frac{1}{\kappa_{s}+1}\mathbf{h}_{r,k}^{H}\mathbf{A}\mathbf{h}_{r,k}\label{eq:54}
\end{equation}

\noindent where

\noindent $\mathbf{A}=\Theta_{s}\left(\kappa_{s}\mathbf{\bar{H}}\mathbf{\bar{H}}^{H}+\sqrt{\kappa_{s}}\mathbf{\bar{H}}\mathbf{\tilde{H}}^{H}+\sqrt{\kappa_{s}}\mathbf{\tilde{H}}\mathbf{\bar{H}}^{H}+\mathbf{\tilde{H}}\mathbf{\tilde{H}}^{H}\right)\Theta_{s}^{H}$.
Now (\ref{eq:54}) can be expressed as $\frac{1}{\kappa_{s}+1}\mathbf{h}_{r,k}^{H}\mathbf{A}\mathbf{h}_{r,k}=\frac{1}{\kappa_{s}+1}\underset{\Delta_{1}}{\underbrace{\mathbf{\bar{h}}_{r,k}^{H}\mathbf{A}\mathbf{\bar{h}}_{r,k}}}$.
After removing the zero expectation terms, we can get
\begin{equation}
\mathscr{E}\left\{ \left|\mathbf{h}_{r,k_{s}}^{H}\Theta_{s}\mathbf{H}_{s}\mathbf{H}_{s}^{H}\Theta_{s}^{H}\mathbf{h}_{r,k_{s}}\right|^{2}\right\} =\frac{1}{\left(\kappa_{s}+1\right)^{2}}\mathscr{E}\left\{ \left|\Delta_{1}\right|^{2}\right\}. 
\end{equation}

\noindent Now we can write the main term as 
\[
\Delta_{1}=\frac{1}{\kappa_{s}+1}\mathbf{\bar{h}}_{r,k}^{H}\Theta_{s}\left(\kappa_{s}\mathbf{\bar{H}}\mathbf{\bar{H}}^{H}+\sqrt{\kappa_{s}}\mathbf{\bar{H}}\mathbf{\tilde{H}}^{H}\right.
\]
\begin{equation}
\left.+\sqrt{\kappa_{s}}\mathbf{\tilde{H}}\mathbf{\bar{H}}^{H}+\mathbf{\tilde{H}}\mathbf{\tilde{H}}^{H}\right)\Theta_{s}^{H}\mathbf{\bar{h}}_{r,k}
\end{equation}
\[
=\left(\underset{\Delta_{1,1}}{\underbrace{\frac{\kappa_{s}}{\kappa_{s}+1}\mathbf{\bar{h}}_{r,k}^{H}\Theta_{s}\mathbf{\bar{H}}\mathbf{\bar{H}}^{H}\Theta_{s}^{H}\mathbf{\bar{h}}_{r,k}}}+\underset{\Delta_{1,2}}{\underbrace{\frac{\sqrt{\kappa_{s}}}{\kappa_{s}+1}\mathbf{\bar{h}}_{r,k}^{H}\Theta_{s}\mathbf{\bar{H}}\mathbf{\tilde{H}}^{H}\Theta_{s}^{H}\mathbf{\bar{h}}_{r,k}}}\right.
\]
\begin{equation}
\left.+\underset{\Delta_{1,3}}{\underbrace{\frac{\sqrt{\kappa_{s}}}{\kappa_{s}+1}\mathbf{\bar{h}}_{r,k}^{H}\Theta_{s}\mathbf{\tilde{H}}\mathbf{\bar{H}}^{H}\Theta_{s}^{H}\mathbf{\bar{h}}_{r,k}}}+\underset{\Delta_{1,4}}{\underbrace{\frac{1}{\kappa_{s}+1}\mathbf{\bar{h}}_{r,k}^{H}\Theta_{s}\mathbf{\tilde{H}}\mathbf{\tilde{H}}^{H}\Theta_{s}^{H}\mathbf{\bar{h}}_{r,k}}}\right).
\end{equation}

\noindent The average can be extended into
\[
\mathscr{E}\left\{ \left|\Delta_{1}\right|^{2}\right\} =\mathscr{E}\left\{ \left|\Delta_{1,1}\right|^{2}\right\} +\mathscr{E}\left\{ \left|\Delta_{1,2}\right|^{2}\right\} +\mathscr{E}\left\{ \left|\Delta_{1,3}\right|^{2}\right\} 
\]
\begin{equation}
+\mathscr{E}\left\{ \left|\Delta_{1,4}\right|^{2}\right\} +2\mathscr{E}\left\{ \Delta_{1,1}\Delta_{1,4}^{H}\right\}. 
\end{equation}
\noindent The average of the first term is
\begin{equation}
\mathscr{E}\left\{ \left|\Delta_{1,1}\right|^{2}\right\} =\left(\frac{\kappa_{s}}{\kappa_{s}+1}\right)^{2}\left(\left|\mathbf{\bar{h}}_{r,k}^{H}\Theta_{s}\mathbf{\bar{H}}\mathbf{\bar{H}}^{H}\Theta_{s}^{H}\mathbf{\bar{h}}_{r,k}\right|^{2}\right).
\end{equation}

\noindent and the average of the second term is
\begin{equation}
\mathscr{E}\left\{ \left|\Delta_{1,2}\right|^{2}\right\} =\frac{\kappa_{s}}{\left(\kappa_{s}+1\right)^{2}}NM\left|f_{k}\right|^{2}
\end{equation}
where, \noindent $f_{k}=\mathbf{a}_{N}\left(\phi_{r}^{a},\phi_{r}^{e}\right)\Theta_{s}\mathbf{\bar{h}}_{r,k}^{H}$.
The average of the third term is
\begin{equation}
\mathscr{E}\left\{ \left|\Delta_{1,3}\right|^{2}\right\} =\frac{\kappa_{s}}{\left(\kappa_{s}+1\right)^{2}}\left(NM\left|f_{k}\right|^{2}\right).
\end{equation}

\noindent The average of the forth term
\begin{equation}
\mathscr{E}\left\{ \left|\Delta_{1,4}\right|^{2}\right\} =\frac{1}{\left(\kappa_{s}+1\right)^{2}}\left(N^{2}M^{2}+NM^{2}\right).
\end{equation}

\noindent The average of the last term is
\begin{equation}
2\mathscr{E}\left\{ \Delta_{1,1}\Delta_{1,4}^{*}\right\} =2\frac{\kappa_{s}}{\left(\kappa_{s}+1\right)^{2}}\mathbf{\bar{h}}_{r,k}^{H}\Theta_{s}\mathbf{\bar{H}}\mathbf{\bar{H}}^{H}\Theta_{s}\mathbf{\bar{h}}_{r,k}\mathbf{\bar{h}}_{r,k}^{H}N\mathbf{I}_{M}\mathbf{\bar{h}}_{r,k}.
\end{equation}

Now, we need to find the expectation of $y$, $\mathscr{E}\left\{ y\right\} =\mathscr{E}\left\{ \frac{1}{\left[\left(\mathbf{G}_{s}\mathbf{G}_{s}^{H}\right)^{-1}\right]_{k_{s},k_{s}}}\right\} $.
Since $\mathbf{G}_{s}\mathbf{G}_{s}^{H}$ has wishart distribution,
we know that
\[
\mathscr{E}\left\{ \mathbf{G}_{s}\mathbf{G}_{s}^{H}\right\} =N\left(\left(\frac{1}{1+\kappa_{s}}\right)\mathbf{H}_{r}^{H}\mathbf{H}_{r}+\right.
\]
\begin{equation}
\left.\left(\frac{\kappa_{s}}{1+\kappa_{s}}\right)\mathbf{H}_{r}^{H}\Theta_{s}\mathbf{a}_{M}^{H}\mathbf{a}_{M}\Theta_{s}^{H}\mathbf{H}_{r}\right).
\end{equation}
Thus the expectation can be found as
\[
\mathscr{E}\left\{ \left(\mathbf{G}_{s}\mathbf{G}_{s}^{H}\right)^{-1}\right\} =\frac{1}{N-K-1}\left(\left(\frac{1}{1+\kappa_{s}}\right)\mathbf{H}_{r}^{H}\mathbf{H}_{r}+\right.
\]
\begin{equation}
\left.\left(\frac{\kappa_{s}}{1+\kappa_{s}}\right)\mathbf{H}_{r}^{H}\Theta_{s}\mathbf{a}_{M}^{H}\mathbf{a}_{M}\Theta_{s}^{H}\mathbf{H}_{r}\right)^{-1}.
\end{equation}

Finally, by substituting $\mathscr{E}\left\{ x\right\} $ and $\mathscr{E}\left\{ y\right\} $
in (\ref{eq:14}) we can obtain the ergodic rate for the common message.
Similarly, by plugging $\mathscr{E}\left\{ y\right\} $ into (\ref{eq:19-1})
we can get the ergodic rate of the private part.

\section*{Appendix C}

\noindent Based on the Woodbury\textquoteright s identity, we have

\[
\left[\left(\varLambda+\delta\mathbf{H}_{r}^{H}\Theta_{s}\mathbf{a}_{M}^{H}\mathbf{a}_{M}\Theta_{s}^{H}\mathbf{H}_{r}\right)^{-1}\right]_{k_{s},k_{s}}=
\]

\begin{equation}
\left[\varLambda^{-1}\right]_{k_{s},k_{s}}-\frac{\delta\left[\varLambda^{-1}\mathbf{H}_{r}^{H}\Theta_{s}\mathbf{a}_{M}^{H}\mathbf{a}_{M}\Theta_{s}^{H}\mathbf{H}_{r}\varLambda^{-1}\right]_{k_{s},k_{s}}}{1+\delta\mathbf{a}_{M}\Theta_{s}^{H}\mathbf{H}_{r}\varLambda^{-1}\mathbf{H}_{r}^{H}\Theta_{s}\mathbf{a}_{M}^{H}}
\end{equation}

\noindent where $\varLambda=\left(\frac{1}{1+\kappa_{s}}\right)\mathbf{H}_{r}^{H}\mathbf{H}_{r}$
and $\delta=\left(\frac{\kappa_{s}}{1+\kappa_{s}}\right)$. Now using
$\Theta_{s}\mathbf{a}_{M}^{H}=\textrm{diag}\left(\mathbf{a}_{M}^{H}\right)\mathbf{\theta_{s}}$,
we can get (\ref{eq:83}),
\begin{figure*}
\[
\left[\varLambda^{-1}\right]_{k_{s},k_{s}}-\frac{\delta\left[\varLambda^{-1}\mathbf{H}_{r}^{H}\textrm{diag}\left(\mathbf{a}_{M}^{H}\right)\mathbf{\theta_{s}}\mathbf{\theta_{s}}^{H}\textrm{diag}\left(\mathbf{a}_{M}\right)\mathbf{H}_{r}\varLambda^{-1}\right]_{k_{s},k_{s}}}{1+\delta\mathbf{\theta_{s}}^{H}\textrm{diag}\left(\mathbf{a}_{M}\right)\mathbf{H}_{r}\varLambda^{-1}\mathbf{H}_{r}^{H}\textrm{diag}\left(\mathbf{a}_{M}^{H}\right)\mathbf{\theta_{s}}}=
\]

\[
\left[\varLambda^{-1}\right]_{k_{s},k_{s}}-\frac{\delta\left[\varLambda^{-1}\mathbf{H}_{r}^{H}\textrm{diag}\left(\mathbf{a}_{M}^{H}\right)\mathbf{\theta_{s}}\mathbf{\theta_{s}}^{H}\textrm{diag}\left(\mathbf{a}_{M}\right)\mathbf{H}_{r}\varLambda^{-1}\right]_{k_{s},k_{s}}}{\frac{1}{M}\mathbf{\theta_{s}}^{H}\mathbf{I}_{M}\mathbf{\theta_{s}}+\delta\mathbf{\theta_{s}}^{H}\textrm{diag}\left(\mathbf{a}_{M}\right)\mathbf{H}_{r}\varLambda^{-1}\mathbf{H}_{r}^{H}\textrm{diag}\left(\mathbf{a}_{M}^{H}\right)\mathbf{\theta_{s}}}=
\]

\begin{equation}
\left[\varLambda^{-1}\right]_{k_{s},k_{s}}-\frac{\delta\left[\varLambda^{-1}\mathbf{H}_{r}^{H}\textrm{diag}\left(\mathbf{a}_{M}^{H}\right)\mathbf{\theta_{s}}\mathbf{\theta_{s}}^{H}\textrm{diag}\left(\mathbf{a}_{M}\right)\mathbf{H}_{r}\varLambda^{-1}\right]_{k_{s},k_{s}}}{\mathbf{\theta_{s}}^{H}\left(\frac{1}{M}\mathbf{I}_{M}+\delta\textrm{diag}\left(\mathbf{a}_{M}\right)\mathbf{H}_{r}\varLambda^{-1}\mathbf{H}_{r}^{H}\textrm{diag}\left(\mathbf{a}_{M}^{H}\right)\right)\mathbf{\theta_{s}}}=\frac{\mathbf{\theta_{s}}^{H}\left\{ \left[\varLambda^{-1}\right]_{k_{s},k_{s}}\mathbf{B}_{ks}-\delta\mathbf{s}\mathbf{s}^{H}\right\} \mathbf{\theta_{s}}}{\mathbf{\theta_{s}}^{H}\mathbf{B}_{ks}\mathbf{\theta_{s}}}=\frac{\mathbf{\theta_{s}}^{H}\mathbf{A}_{ks}\mathbf{\theta_{s}}}{\mathbf{\theta_{s}}^{H}\mathbf{B}_{ks}\mathbf{\theta_{s}}}\label{eq:83}
\end{equation}

\rule[1ex]{2\columnwidth}{1pt}

\end{figure*}
 where $\mathbf{B}_{ks}=\left(\frac{1}{M}\mathbf{I}_{M}+\delta\textrm{diag}\left(\mathbf{a}_{M}\right)\mathbf{H}_{r}\varLambda^{-1}\mathbf{H}_{r}^{H}\textrm{diag}\left(\mathbf{a}_{M}^{H}\right)\right)$,
$\mathbf{s}=\left[\varLambda^{-1}\mathbf{H}_{r}^{H}\textrm{diag}\left(\mathbf{a}_{M}^{H}\right)\right]_{\left(:,k\right)}$,
$\mathbf{s}^{H}=\left[\varLambda^{-1}\mathbf{H}_{r}^{H}\textrm{diag}\left(\mathbf{a}_{M}^{H}\right)\right]_{\left(k,:\right)}$
and $\mathbf{A}_{ks}=\left\{ \left[\varLambda^{-1}\right]_{k_{s},k_{s}}\mathbf{B}_{ks}-\delta\mathbf{s}\mathbf{s}^{H}\right\} $.

\bibliographystyle{IEEEtran}
\bibliography{bib}

\end{document}